\documentclass[10pt]{article}
\usepackage{graphicx}
\usepackage{amsmath}
\usepackage{amssymb}
\usepackage{caption2}
\begin{document}
\begin{center}
{\large {\bf \sc{  Vector tetraquark state candidates: $Y(4260/4220)$, $Y(4360/4320)$, $Y(4390)$ and $Y(4660/4630)$   }}} \\[2mm]
Zhi-Gang  Wang \footnote{E-mail: zgwang@aliyun.com.  }     \\
 Department of Physics, North China Electric Power University, Baoding 071003, P. R. China
\end{center}

\begin{abstract}
In this article, we construct the $C \otimes \gamma_\mu C$ and $C\gamma_5 \otimes \gamma_5\gamma_\mu C$ type  currents to interpolate  the vector   tetraquark states,  then carry out the operator product expansion up to the vacuum condensates of dimension-10  in   a consistent way, and obtain four QCD sum rules. In calculations,  we use the  formula $\mu=\sqrt{M^2_{Y}-(2{\mathbb{M}}_c)^2}$ to determine  the optimal energy scales of the QCD spectral densities, moreover, we take the experimental values of the masses of the $Y(4260/4220)$, $Y(4360/4320)$, $Y(4390)$ and $Y(4660/4630)$ as input parameters and  fit the pole residues to reproduce the correlation functions at the QCD side. The numerical results support assigning  the $Y(4660/4630)$  to be the  $C \otimes \gamma_\mu C$ type vector tetraquark state $c\bar{c}s\bar{s}$, assigning the $Y(4360/4320)$ to be $C\gamma_5 \otimes \gamma_5\gamma_\mu C$  type vector tetraquark state $c\bar{c}q\bar{q}$,  and disfavor assigning the $Y(4260/4220)$ and $Y(4390)$ to be the pure vector tetraquark states.
 \end{abstract}



\section{Introduction}

In 2005, the BaBar collaboration  studied the initial-state radiation process  $e^+ e^- \to \gamma_{ISR} \pi^+\pi^- J/\psi$ and observed the $Y(4260)$   in the $\pi^+\pi^- J/\psi$ invariant-mass spectrum  \cite{BaBar4260-0506}. Later the $Y(4260)$ was confirmed by the Belle and CLEO collaborations \cite{Belle-0707,CLEO-0606},
the Belle collaboration also observed an evidence for a very broad structure $Y(4008)$ in the $\pi^+\pi^- J/\psi$ mass spectrum.
In 2014, the BES collaboration searched for the production of $e^+e^-\to \omega\chi_{cJ}$ with $J=0,1,2$, and observed a resonance in the $\omega\chi_{c0}$ cross section,  the measured mass and width of the resonance  are $4230\pm 8\pm 6\, \rm{ MeV}$    and $ 38\pm 12\pm 2\,\rm{MeV}$, respectively \cite{BES-2014-4230}.

In 2016, the BES collaboration measured the cross sections of the process $e^+ e^- \to \pi^+\pi^- h_c$, and observed two structures, the $Y(4220)$ has a mass of $4218.4\pm4.0\pm0.9\,\rm{MeV}$ and a width of $66.0\pm9.0\pm0.4\,\rm{MeV}$ respectively, and the $Y(4390)$ has a mass  of $4391.6\pm6.3\pm1.0\,\rm{MeV}$ and a width of $139.5\pm16.1\pm0.6\,\rm{MeV}$ respectively \cite{BES-Y4390}.

Also in 2016, the BES collaboration precisely measured the cross section of the process $e^+ e^- \to  \pi^+\pi^- J/\psi$   at center-of-mass energies from $3.77$ to $4.60\,\rm{GeV}$  and observed  two resonant structures. The first resonance has a mass of $4222.0\pm3.1\pm 1.4\,  \rm{MeV}$ and a width of $44.1 \pm 4.3\pm 2.0 \,\rm{MeV}$,  the second one has a mass of $4320.0\pm 10.4 \pm 7.0\, \rm{MeV}$   and a width of $101.4^{+25.3}_{-19.7}\pm 10.2\,\rm{MeV}$ \cite{BES-Y4220-Y4320}.
The  first resonance agrees with the $Y(4260)$ while the second resonance agrees with the $Y(4360)$ according to the uncertainties, the $Y(4008)$ resonance previously observed by the Belle experiment is not confirmed \cite{Belle-0707}.

In Ref.\cite{CPShen},  Gao,  Shen and Yuan perform a combined fit for the cross sections of $e^+e^-\to \omega \chi_{c0}$, $\pi^+\pi^-h_c$, $\pi^+\pi^- J/\psi$, $D^0 D^{*-}\pi^+ + c.c.$ measured by the BESIII experiment, and determine   a mass $4219.6\pm 3.3\pm 5.1 \, \rm{MeV}$  and a total width $56.0\pm 3.6\pm 6.9 \, \rm{MeV}$ for the $Y(4220)$.

In 2006, the BaBar collaboration observed an evidence for a broad structure at  $4.32\,\rm{GeV}$ in the $\pi^+\pi^- \psi^\prime$ mass spectrum  in the process $e^+e^- \to \pi^+ \pi^- \psi^{\prime}$ \cite{BaBar-Y4360}.
In 2007, the  Belle collaboration  studied the initial-state radiation process $e^+e^- \to \gamma_{ISR}\pi^+ \pi^- \psi^{\prime}$, and  observed two structures $Y(4360)$ and $Y(4660)$ in the $\pi^+ \pi^- \psi^{\prime}$ invariant-mass spectrum \cite{Belle4660-0707}.
In 2008, the Belle collaboration studied  the initial-state radiation process $e^+e^- \to \gamma_{ISR} \Lambda_c^+ \Lambda_c^-$   and observed a clear peak $Y(4630)$   in the $\Lambda_c^+ \Lambda_c^-$  invariant-mass spectrum \cite{Belle4630-0807}. The $Y(4360)$ and $Y(4660/4630)$ were confirmed by the BaBar collaboration \cite{BaBar-Y4360-Y4660}.

There have been several assignments for those $Y$ states, such as the tetraquark states \cite{Maiani-4260,Maiani-II-type,Tetraquark-PRT,Ali-Maiani-Y,Brodsky-PRL,WangJPG,ZhangHuang-PRD,ZhangHuang-JHEP,Nielsen-4260-4460,WangEPJC-4660,WangEPJC-1601}, hybrid states \cite{Hybrid-4260,Hybrid-4260-Lattice,BO-potential}, hadro-charmonium states \cite{Voloshin-Y4260,Voloshin-More,GuoFK-4660-psif0,WangZG-4660-psif0},
 molecular states \cite{Zhao-PRL-Y4260-3900,Zhao-PRL-Y4260-3900-2,Wang-Y4260-No-hadro,WangCPC-Y4390},  kinematical effects \cite{Chen-He-Liu-4260,CC-Effects}, baryonium states \cite{Qiao-CF-4260}, etc. The $Y(4260)$, which is the milestone of the $Y$ states,  has been extensively studied.

In Ref.\cite{Maiani-4260}, L. Maiani et al assign the $Y(4260)$ to be the first orbital excitation
of a diquark-antidiquark state $[cs]_{S=0}[\bar{c}\bar{s}]_{S=0}$  based on the effective  Hamiltonian with the spin-spin and spin-orbit  interactions \cite{Maiani-3872}, where the subscript $S$ denotes the diquark spins. In the type-II diquark-antidiquark model \cite{Maiani-II-type}, where the spin-spin interactions between the quarks and antiquarks are neglected, L. Maiani et al interpret the $Y(4360)$ and $Y(4660)$ as the first radial excitations of the $Y(4008)$ and $Y(4260)$ respectively,
and interpret the $Y(4008)$, $Y(4260)$, $Y(4290/4220)$  and $Y(4630)$ as the four ground  states with the angular momentum $L=1$. One can consult Ref.\cite{Tetraquark-PRT} for detailed reviews of the effective  Hamiltonian approach.

In Ref.\cite{Ali-Maiani-Y}, A. Ali et al analyze the hidden-charm P-wave tetraquarks and the newly excited charmed $\Omega_c$ states in the diquark model using the  effective Hamiltonian incorporating the dominant spin-spin, spin-orbit and tensor interactions,  and observe that the preferred  assignments of the ground state tetraquark states with $L=1$ are the $Y(4220)$, $Y(4330)$, $Y(4390)$, $Y(4660)$ rather than the  $Y(4008)$, $Y(4260)$, $Y(4360)$, $Y(4660)$.

In the effective Hamiltonian,  an approximately common rest frame for all the components is assumed, while in the dynamical diquark picture,
 the diquark-antidiquark pair forms promptly at the production point, and rapidly
separates due to the kinematics of the production process, then they create a color flux tube or string between them \cite{Brodsky-PRL}.

Another possible assignment of the $Y(4260)$ is the $c\bar{c}g$ hybrid state \cite{Hybrid-4260}, the lattice calculations indicate that  the vector charmonium hybrid has a mass about $4285\,\rm{MeV}$, which is quite close to that of the $Y(4260)$ \cite{Hybrid-4260-Lattice}. Furthermore,  lattice results strongly indicate that the quarks
should form a spin singlet in the low-lying vector hybrid, which in rough agreement with the spin-flip suppression  in the  annihilation $e^+e^-\to Y(4260)$.
An alternative approach is to describe the $X$, $Y$,  $Z$ mesons  as bound states in the  Born-Oppenheimer  potentials (usually lattice-QCD computed gluon-induced potentials) for a heavy quark and a heavy antiquark \cite{BO-potential}. For more literatures on the $Q\bar{Q}$ potentials involving or not involving of the gluonic excitations, one can consult Refs.\cite{pNRQCD,HeavyQuarkonium-EPJC}.

In Ref.\cite{Voloshin-Y4260}, Li and Voloshin  suggest that the $Y(4260)$ and $Y(4360)$ are a mixture, with mixing close to maximal, of two states of the hadro-charmonium, one containing a spin-triplet  $c\bar{c}$   pair and the other containing a spin-singlet $c\bar{c}$ pair. While the $Y(4660)$ can be assigned to be the $\psi^\prime f_0(980)$ hadro-charmonium or molecular   state (the two scenarios overlap in this case) \cite{GuoFK-4660-psif0,WangZG-4660-psif0}. For more literatures on the hadro-charmonium states, one can consult Ref.\cite{Voloshin-More}.

In the   scenario of molecular   states, the $Y(4260)$ and $Z_c(3900)$ are assigned to be the $\bar{D}D_1(2420) +D\bar{D}_1(2420)$ and $\bar D D^*+D\bar D^*$ molecular states respectively  \cite{Zhao-PRL-Y4260-3900,Zhao-PRL-Y4260-3900-2}, which
are compatible  with the processes   $e^+e^-\to J/\psi\pi^+\pi^-\, ,\, h_c\pi^+\pi^-$ measured by the BESIII and Belle collaborations.
In Ref.\cite{Wang-Y4260-No-hadro}, Q. Wang et al confront both the hadronic molecule and the hadro-charmonium interpretations
of the $Y(4260)$ with the available experimental data, and conclude that the data support the $Y(4260)$ being dominantly a $\bar{D}D_1(2420) +D\bar{D}_1(2420)$ hadronic molecule while they challenge the hadro-charmonium interpretation.
The assignment of the $Y(4260)$ as the tetraquark state or molecular state means different decay ratio $Y(4260)\to \gamma\,X(3872)$ \cite{Maiani-gamma,Guo-gamma}.
Precisely measuring the decay ratio can shed light on the  nature of the $Y(4260)$, $Z_c(3900)$ and $X(3872)$.

In other interpretations, the three charmonium-like states $Y(4008)$, $Y(4260)$  and $Y(4360)$  are Fango-like interference
phenomena or coupled-channel effects rather than  genuine resonances \cite{Chen-He-Liu-4260,CC-Effects}.  While C. F. Qiao assigns  the $Y(4260)$ to be a baryonium state $\Lambda_c^+\Lambda_c^-$ \cite{Qiao-CF-4260}.

In this article, we tentatively assume that there exist four exotic vector  $Y$ states,  $Y(4260/4220)$, $Y(4360/4320)$, $Y(4390)$ and $Y(4660/4630)$,
\begin{eqnarray}
Y(4220)&\to& \omega\chi_{c0}\, , \, J/\psi\pi^+\pi^-\, , \,  h_c\pi^+\pi^-   \,\,\, \cite{BES-2014-4230,BES-Y4390,BES-Y4220-Y4320}\, ,\nonumber \\
Y(4320)&\to& J/\psi\pi^+\pi^-\, , \,\psi^\prime \pi^+\pi^-   \,\,\,  \cite{BES-Y4220-Y4320,BaBar-Y4360,Belle4660-0707}\, ,\nonumber \\
Y(4390)&\to& h_c\pi^+\pi^-   \,\,\, \cite{BES-Y4390}\, ,\nonumber \\
Y(4660)&\to& \psi^\prime \pi^+\pi^- \, ,\, \Lambda_c^+ \Lambda_c^-   \,\,\, \cite{Belle4660-0707,Belle4630-0807} \, ,
\end{eqnarray}
and  will focus on  the  scenario of tetraquark  states based on the QCD sum rules.

The diquarks $\varepsilon^{ijk}q^{T}_j C\Gamma q^{\prime}_k$ have  five  structures  in Dirac spinor space, where $C\Gamma=C\gamma_5$, $C$, $C\gamma_\mu \gamma_5$,  $C\gamma_\mu $ and $C\sigma_{\mu\nu}$ for the scalar, pseudoscalar, vector, axialvector  and  tensor diquarks, respectively.
The attractive interactions of one-gluon exchange  favor  formation of
the diquarks in  color antitriplet, flavor
antitriplet and spin singlet \cite{One-gluon},
 while the favored configurations are the scalar ($C\gamma_5$) and axialvector ($C\gamma_\mu$) diquark states based on the QCD sum rules \cite{WangDiquark,WangLDiquark,Tang-Diquark,Dosch-Diquark-1989}.
 The $C\gamma_5$ type and $C\gamma_\mu$ type diquark states can be reduced to the non-relativistic spin-0 and spin-1 diquark states respectively in the  effective Hamiltonian approach \cite{Tetraquark-PRT}.

We can construct the  diquark-antidiquark type  currents
\begin{eqnarray}
&&C\gamma_5 \otimes \gamma_5C\, , \nonumber\\
&&C\gamma_\mu \otimes \gamma^\mu C\, ,
\end{eqnarray}
 to interpolate the scalar  hidden-charm tetraquark states \cite{WangScalarT,WangMPLA,WangTetraquarkCTP} or construct the  diquark-antidiquark type currents
\begin{eqnarray}
&&C\otimes \gamma_\mu C\, , \nonumber\\
&&C\gamma_5 \otimes \gamma_5\gamma_\mu C\, ,\nonumber\\
&&C\gamma_5\otimes\partial_\mu \gamma_5C\, , \nonumber \\
&&C\gamma_\alpha \otimes \partial_\mu \gamma^\alpha C\, , \nonumber \\
&&C\gamma_\mu \otimes \gamma_\nu C-C\gamma_\nu \otimes \gamma_\mu C\, ,
\end{eqnarray}
to interpolate the vector hidden-charm tetraquark states \cite{WangJPG,ZhangHuang-PRD,ZhangHuang-JHEP,Nielsen-4260-4460,WangEPJC-4660,WangEPJC-1601}.
One can consult Ref.\cite{ChenZhu} for more interpolating currents for the vector tetraquark states without introducing an additional P-wave between the diquark and antidiquark. We can also
choose the mixed charmonium-tetraquark currents to study the vector mesons $Y(4260)$ and $Y(4360)$ \cite{WangEPJC-1601,Nielsen4260-1209}.

\begin{table}
\begin{center}
\begin{tabular}{|c|c|c|c|c|c|c|c|}\hline\hline
           &Structures                                    &Constituents                           &OPE\,(No)    &mass(GeV)      &References   \\ \hline
$Y(4660)$  &$C\gamma_5\otimes\partial_\mu \gamma_5C$      &$c\bar{c}s\bar{s}$                     &$6$          &$4.69$         &\cite{ZhangHuang-PRD}  \\ \hline

$Y(4660)$  &$C\gamma_5 \otimes \gamma_5\gamma_\mu C$      &$c\bar{c}s\bar{s}$                     &$8\,(7)$     &$4.65$         &\cite{Nielsen-4260-4460}  \\ \hline

$Y(4660)$  &$C \otimes \gamma_\mu C$                      &$c\bar{c}s\bar{s}/c\bar{c}q\bar{q}$    &$10$         &$4.70/4.66$    &\cite{WangEPJC-4660}  \\ \hline

$Y(4660)$  &$C\gamma_\mu \otimes \gamma_\nu C-C\gamma_\nu \otimes \gamma_\mu C$   &$c\bar{c}q\bar{q}$  &$10$    &$4.66$         &\cite{WangEPJC-1601}  \\ \hline

$Y(4660)$  &$C\gamma_5 \otimes \gamma_5\gamma_\mu C$      &$c\bar{c}q\bar{q}$                     &$8\,(7)$     &$4.64$         &\cite{ChenZhu}  \\ \hline

$Y(4360)$  &$C\gamma_5\otimes\partial_\mu \gamma_5C$      &$c\bar{c}q\bar{q}$                     &$6$          &$4.32$         &\cite{ZhangHuang-JHEP}  \\ \hline

$Y(4360)$  &$C \otimes \gamma_\mu C\oplus \gamma_\mu$     &$c\bar{c}q\bar{q}\oplus c\bar{c}$      &$10$         &$4.36$         &\cite{WangEPJC-1601}  \\ \hline

$Y(4260)$  &$C \otimes \gamma_\mu C\oplus \gamma_\mu$     &$c\bar{c}q\bar{q}\oplus c\bar{c}$      &$10$         &$4.26$         &\cite{WangEPJC-1601}  \\ \hline

$Y(4260)$  &$C\gamma_5 \otimes \gamma_5\gamma_\mu C\oplus \gamma_\mu$  &$c\bar{c}q\bar{q}\oplus c\bar{c}$  &$8\,(7)$  &$4.26$   &\cite{Nielsen4260-1209}  \\ \hline \hline
\end{tabular}
\end{center}
\caption{ The OPE denotes  truncations of  the operator product expansion up to the vacuum condensates of dimension $n$, the No denotes the vacuum condensates of dimension $n^\prime$ are not included.   }
\end{table}

In Table 1, we present the existing predictions of the masses of the vector tetraquark states based on the QCD sum rules. In
 Refs.\cite{ZhangHuang-PRD,ZhangHuang-JHEP,Nielsen-4260-4460,ChenZhu,Nielsen4260-1209}, the vacuum condensates are taken at the energy scale $\mu=1\,\rm{GeV}$ while the $\overline{MS}$ mass $m_c(m_c)$ is taken at the energy scale $\mu=m_c(m_c)$, the energy scales of the QCD spectral densities are not specified.
In Refs.\cite{WangEPJC-4660,WangEPJC-1601,WangTetraquarkCTP}, we take the formula $\mu=\sqrt{M^2_{X/Y/Z}-(2{\mathbb{M}}_{c})^2}$   with the effective charmed  quark mass ${\mathbb{M}}_{c}$  to determine  the energy scales of the QCD spectral densities of the hidden-charm  tetraquark states, and evolve  the vacuum condensates and the $\overline{MS}$ mass $m_c(m_c)$  to the optimal energy scales $\mu$ to extract the tetraquark masses;  the hidden-bottom tetraquark states can be studied analogously  \cite{WangHuangtao-2014}.

In this article, we assume that the $Y(4260/4220)$, $Y(4360/4320)$, $Y(4390)$ and $Y(4660/4630)$ are vector tetraquark states, and restudy the $C \otimes \gamma_\mu C$ and $C\gamma_5 \otimes \gamma_5\gamma_\mu C$  type vector tetraquark states with the QCD sum rules in details  by taking into account the vacuum condensates up to dimension 10 in a consistent way in the operator product expansion, and use the energy scale formula $\mu=\sqrt{M^2_{X/Y/Z}-(2{\mathbb{M}}_c)^2}$  to determine the optimal energy scales of the QCD spectral densities. In the QCD sum rules for the tetraquark states, the terms associate with $\frac{1}{T^2}$, $\frac{1}{T^4}$, $\frac{1}{T^6}$ in the QCD spectral densities manifest themselves at small values of the Borel parameter $T^2$, we have to choose large values of the $T^2$ to warrant convergence of the operator product expansion and appearance of the Borel platforms. The higher dimensional  vacuum condensates play an important role in determining the Borel windows therefore the ground state  masses and pole residues, though they maybe play a less important role in the Borel windows. We should take them into account consistently.

The article is arranged as follows:  we derive the QCD sum rules for the masses and pole residues of  the vector   tetraquark states in section 2; in section 3, we   present the numerical results and discussions; section 4 is reserved for our conclusion.

\section{QCD sum rules for  the  vector tetraquark states}
In the following, we write down  the two-point correlation functions $\Pi_{\mu\nu}(p)$  in the QCD sum rules,
\begin{eqnarray}
\Pi_{\mu\nu}(p)&=&i\int d^4x e^{ip \cdot x} \langle0|T\left\{J_\mu(x)J_\nu^{\dagger}(0)\right\}|0\rangle \, , \\
J^1_\mu(x)&=&\frac{\varepsilon^{ijk}\varepsilon^{imn}}{\sqrt{2}}\left\{s^{Tj}(x)C c^k(x) \bar{s}^m(x)\gamma_\mu C \bar{c}^{Tn}(x)-s^{Tj}(x)C\gamma_\mu c^k(x)\bar{s}^m(x)C \bar{c}^{Tn}(x) \right\} \, ,\nonumber \\
J^2_\mu(x)&=&\frac{\varepsilon^{ijk}\varepsilon^{imn}}{2}\left\{u^{Tj}(x)C c^k(x) \bar{u}^m(x)\gamma_\mu C \bar{c}^{Tn}(x)+d^{Tj}(x)C c^k(x) \bar{d}^m(x)\gamma_\mu C \bar{c}^{Tn}(x) \right.\nonumber\\
&&\left.-u^{Tj}(x)C\gamma_\mu c^k(x)\bar{u}^m(x)C \bar{c}^{Tn}(x) -d^{Tj}(x)C\gamma_\mu c^k(x)\bar{d}^m(x)C \bar{c}^{Tn}(x)\right\} \, , \\
J^3_\mu(x)&=&\frac{\varepsilon^{ijk}\varepsilon^{imn}}{\sqrt{2}}\left\{s^{Tj}(x)C\gamma_5 c^k(x) \bar{s}^m(x)\gamma_5\gamma_\mu C \bar{c}^{Tn}(x)+s^{Tj}(x)C\gamma_\mu \gamma_5 c^k(x)\bar{s}^m(x)\gamma_5 C \bar{c}^{Tn}(x) \right\} \, , \nonumber \\
J^4_\mu(x)&=&\frac{\varepsilon^{ijk}\varepsilon^{imn}}{2}\left\{u^{Tj}(x)C\gamma_5 c^k(x) \bar{u}^m(x)\gamma_5\gamma_\mu C \bar{c}^{Tn}(x)+d^{Tj}(x)C\gamma_5 c^k(x) \bar{d}^m(x)\gamma_5\gamma_\mu C \bar{c}^{Tn}(x) \right.\nonumber\\
&&\left.+u^{Tj}(x)C\gamma_\mu\gamma_5 c^k(x)\bar{u}^m(x)\gamma_5C \bar{c}^{Tn}(x) +d^{Tj}(x)C\gamma_\mu\gamma_5 c^k(x)\bar{d}^m(x)\gamma_5C \bar{c}^{Tn}(x)\right\} \, ,
\end{eqnarray}
where $\Pi_{\mu\nu}(p)=\Pi^1_{\mu\nu}(p)$, $\Pi^2_{\mu\nu}(p)$, $\Pi^3_{\mu\nu}(p)$, $\Pi^4_{\mu\nu}(p)$, $J_\mu(x)=J_\mu^1(x)$, $J_\mu^2(x)$, $J_\mu^3(x)$, $J_\mu^4(x)$,   the $i$, $j$, $k$, $m$, $n$ are color indexes, the $C$ is the charge conjugation matrix.
 Under charge conjugation transform $\widehat{C}$, the currents $J_\mu(x)$ have the properties,
\begin{eqnarray}
\widehat{C}J_{\mu}(x)\widehat{C}^{-1}&=&- J_\mu(x) \, .
\end{eqnarray}

In the non-relativistic diquark-antidiquark model, one often introduces  an explicit P-wave between the diquark and antidiquark  in the ground state
$C\gamma_5 \otimes \gamma_5C$ type or $C\gamma_5 \otimes \gamma_\mu C$ type or $C\gamma_\mu \otimes \gamma_\nu C$ type tetraquark state \cite{Maiani-4260,Maiani-II-type,Tetraquark-PRT,Ali-Maiani-Y}.  In this article, we choose the $C\otimes \gamma_\mu C$ type and
$C\gamma_5 \otimes \gamma_5\gamma_\mu C$ type vector currents to interpolate the vector tetraquark states, the net effects of the relative P-waves   are embodied in the underlined  $\gamma_5$ in the  $C\gamma_5 \underline{\gamma_5} \otimes \gamma_\mu C$ type and
$C\gamma_5 \otimes \underline{\gamma_5}\gamma_\mu C$ type currents or in the underlined  $\gamma^\alpha$ in the  $C\gamma_\alpha \underline{\gamma^\alpha} \otimes \gamma_\mu C$ type  currents.

The tetraquark states are spatial extended objects, not point-like objects \cite{Brodsky-PRL}, in the QCD sum rules \cite{WangJPG,ZhangHuang-PRD,ZhangHuang-JHEP,Nielsen-4260-4460,WangEPJC-4660,WangEPJC-1601} and the effective Hamiltonian approach \cite{Maiani-4260,Maiani-II-type,Tetraquark-PRT,Ali-Maiani-Y}, the finite size effects are neglected, which leads to some uncertainties, while in the potential models,  an explicit spatial extended potential between the diquark and antiquark is introduced \cite{Ebert-potential}.

Now we perform Fierz re-arrangement  to the vector currents $J_{\mu}^{1}(x)$ and $J_{\mu}^{3}(x)$ both in the color and Dirac-spinor  spaces,  and obtain the following results,
\begin{eqnarray}
J_{\mu}^{1} &=&\frac{1}{2\sqrt{2}}\Big\{\,\bar{c} \gamma^\mu c\,\bar{s} s-\bar{c} c\,\bar{s}\gamma^\mu s+i\bar{c}\gamma^\mu\gamma_5 s\,\bar{s}i\gamma_5 c-i\bar{c} i\gamma_5 s\,\bar{s}\gamma^\mu \gamma_5c  \nonumber\\
&&  - i\bar{c}\gamma_\nu\gamma_5c\, \bar{s}\sigma^{\mu\nu}\gamma_5s+i\bar{c}\sigma^{\mu\nu}\gamma_5c\, \bar{s}\gamma_\nu\gamma_5s
- i\bar{s}\gamma_\nu c\, \bar{c}\sigma^{\mu\nu}s+i \bar{s}\sigma^{\mu\nu}c \,\bar{c}\gamma_\nu s  \,\Big\} \, , \\
J^3_\mu &=&\frac{1}{2\sqrt{2}}\Big\{\,\bar{c}c\,\bar{s}\gamma^\mu s+\bar{c} \gamma^\mu c\,\bar{s} s-\bar{c}\gamma^\mu s\,\bar{s} c-\bar{c} s\,\bar{s}\gamma^\mu c -i\bar{c}\sigma^{\mu\nu} \gamma_5c\, \bar{s}\gamma_\nu \gamma_5s\nonumber\\
&& - i\bar{c}\gamma_\nu \gamma_5c\, \bar{s}\sigma^{\mu\nu} \gamma_5s
+ i\bar{s}\gamma_\nu \gamma_5c\, \bar{c}\sigma^{\mu\nu}\gamma_5s+i \bar{s}\sigma^{\mu\nu} \gamma_5c \,\bar{c}\gamma_\nu\gamma_5 s  \,\Big\} \, ,
\end{eqnarray}
the Fierz re-arrangement of the $J_{\mu}^{2}(x)$ and $J_{\mu}^{4}(x)$ can be obtained analogously.
 The
diquark-antidiquark type current with special quantum numbers couples potentially  to a special tetraquark
state, while the current can be re-arranged to a current as a special superposition of color singlet-singlet type currents, which  couple potentially
 to the meson-meson pairs or molecular states. The diquark-antidiquark type tetraquark
state can be taken as a special superposition of a series of meson-meson pairs, and embodies
the net effects.

According to the current-meson couplings,
\begin{eqnarray}
\langle 0|\bar{c}(0) c(0)|\chi_{c0}(p)\rangle &=&f_{\chi_{c0}}\,M_{\chi_{c0}}\, ,\nonumber\\
\langle 0|\bar{c}(0)\gamma_{\mu} c(0)|J/\psi(p)\rangle &=&f_{J/\psi}\, M_{J/\psi}\,e_{\mu}\, , \nonumber\\
\langle 0|\bar{c}(0)\sigma_{\mu\nu}\gamma_5 c(0)|h_c(p)\rangle &=&if_{h_c}\, \left(e_\mu p_\nu-e_\nu p_\mu \right)\, ,\nonumber\\
\langle 0|\bar{c}(0)\sigma_{\mu\nu}\gamma_5 c(0)|J/\psi(p)\rangle &=&if^T_{J/\psi}\, \varepsilon_{\mu\nu\alpha\beta}\,e^\alpha p^\beta\, ,
\end{eqnarray}
where the $e_\mu$ are the polarization vectors of the $J/\psi$ and $h_c$, we can obtain the conclusion, if the  $Y(4260/4220)$, $Y(4360/4320)$, $Y(4390)$ and $Y(4660/4630)$ are the $C\otimes \gamma_\mu C$ type or $C\gamma_5 \otimes \gamma_5\gamma_\mu C$ type tetraquark states, there are no  heavy quark spin-flips  in the decays to the final  states  $J/\psi$ and $h_c$, which is consistent with the experimental data \cite{BES-2014-4230,BES-Y4390,BES-Y4220-Y4320,BaBar-Y4360,Belle4660-0707,Belle4630-0807}.

At the hadronic side, we can insert  a complete set of intermediate hadronic states with
the same quantum numbers as the current operators $J_\mu(x)$ into the
correlation functions $\Pi_{\mu\nu}(p)$  to obtain the hadronic representation
\cite{SVZ79,Reinders85}. After isolating the ground state
contributions of the vector tetraquark states which are supposed to be the $Y(4260/4220)$, $Y(4360/4320)$, $Y(4390)$ and $Y(4660/4630)$, we get the following results,
\begin{eqnarray}
\Pi_{\mu\nu}(p)&=&\frac{\lambda_{Y}^2}{M_{Y}^2-p^2}\left(-g_{\mu\nu} +\frac{p_\mu p_\nu}{p^2}\right) +\cdots \, \, ,\nonumber\\
&=&\Pi(p^2)\left(-g_{\mu\nu} +\frac{p_\mu p_\nu}{p^2}\right) +\cdots \, ,
\end{eqnarray}
where the pole residues  $\lambda_{Y}$ are defined by
\begin{eqnarray}
 \langle 0|J_\mu(0)|Y(p)\rangle=\lambda_{Y} \,\varepsilon_\mu \, ,
\end{eqnarray}
the $\varepsilon_\mu$ are the polarization vectors of the  vector tetraquark states $Y(4260/4220)$, $Y(4360/4320)$, $Y(4390)$ and $Y(4660/4630)$, etc.

 In the following,  we take the currents $J_\mu^1(x)$ and $J^3_\mu(x)$ as an example, and briefly outline  the operator product expansion for the correlation functions $\Pi_{\mu\nu}(p)$  in perturbative QCD.  We contract the $c$ and $s$ quark fields in the correlation functions
$\Pi_{\mu\nu}(p)$ with Wick theorem, obtain the results:
\begin{eqnarray}
\Pi_{\mu\nu}^{1}(p)&=&\frac{i\varepsilon^{ijk}\varepsilon^{imn}\varepsilon^{i^{\prime}j^{\prime}k^{\prime}}\varepsilon^{i^{\prime}m^{\prime}n^{\prime}}}{2}\int d^4x e^{ip \cdot x}   \nonumber\\
&&\left\{{\rm Tr}\left[ C^{kk^{\prime}}(x) CS^{jj^{\prime}T}(x)C\right] {\rm Tr}\left[ \gamma_\nu C^{n^{\prime}n}(-x)\gamma_\mu C S^{m^{\prime}mT}(-x)C\right] \right. \nonumber\\
&&+{\rm Tr}\left[ \gamma_\mu C^{kk^{\prime}}(x)\gamma_\nu CS^{jj^{\prime}T}(x)C\right] {\rm Tr}\left[  C^{n^{\prime}n}(-x) C S^{m^{\prime}mT}(-x)C\right] \nonumber\\
&&+{\rm Tr}\left[ \gamma_\mu C^{kk^{\prime}}(x) CS^{jj^{\prime}T}(x)C\right] {\rm Tr}\left[ \gamma_\nu C^{n^{\prime}n}(-x) C S^{m^{\prime}mT}(-x)C\right] \nonumber\\
 &&\left.+{\rm Tr}\left[  C^{kk^{\prime}}(x)\gamma_\nu CS^{jj^{\prime}T}(x)C\right] {\rm Tr}\left[  C^{n^{\prime}n}(-x)\gamma_\mu C S^{m^{\prime}mT}(-x)C\right] \right\} \, ,
\end{eqnarray}

\begin{eqnarray}
\Pi_{\mu\nu}^{3}(p)&=&\frac{i\varepsilon^{ijk}\varepsilon^{imn}\varepsilon^{i^{\prime}j^{\prime}k^{\prime}}\varepsilon^{i^{\prime}m^{\prime}n^{\prime}}}{2}\int d^4x e^{ip \cdot x}   \nonumber\\
&&\left\{{\rm Tr}\left[\gamma_5 C^{kk^{\prime}}(x)\gamma_5 CS^{jj^{\prime}T}(x)C\right] {\rm Tr}\left[\gamma_5 \gamma_\nu C^{n^{\prime}n}(-x)\gamma_\mu \gamma_5C S^{m^{\prime}mT}(-x)C\right] \right. \nonumber\\
&&+{\rm Tr}\left[ \gamma_\mu \gamma_5C^{kk^{\prime}}(x)\gamma_5\gamma_\nu CS^{jj^{\prime}T}(x)C\right] {\rm Tr}\left[ \gamma_5 C^{n^{\prime}n}(-x) \gamma_5C S^{m^{\prime}mT}(-x)C\right] \nonumber\\
&&-{\rm Tr}\left[ \gamma_\mu\gamma_5 C^{kk^{\prime}}(x)\gamma_5 CS^{jj^{\prime}T}(x)C\right] {\rm Tr}\left[ \gamma_5\gamma_\nu C^{n^{\prime}n}(-x) \gamma_5C S^{m^{\prime}mT}(-x)C\right] \nonumber\\
 &&\left.-{\rm Tr}\left[  \gamma_5C^{kk^{\prime}}(x)\gamma_5\gamma_\nu CS^{jj^{\prime}T}(x)C\right] {\rm Tr}\left[ \gamma_5 C^{n^{\prime}n}(-x)\gamma_\mu \gamma_5C S^{m^{\prime}mT}(-x)C\right] \right\} \, ,
\end{eqnarray}
where  the $S_{ij}(x)$ and $C_{ij}(x)$ are the full $s$ and $c$ quark propagators respectively,
 \begin{eqnarray}
S_{ij}(x)&=& \frac{i\delta_{ij}\!\not\!{x}}{ 2\pi^2x^4}
-\frac{\delta_{ij}m_s}{4\pi^2x^2}-\frac{\delta_{ij}\langle
\bar{s}s\rangle}{12} +\frac{i\delta_{ij}\!\not\!{x}m_s
\langle\bar{s}s\rangle}{48}-\frac{\delta_{ij}x^2\langle \bar{s}g_s\sigma Gs\rangle}{192}+\frac{i\delta_{ij}x^2\!\not\!{x} m_s\langle \bar{s}g_s\sigma
 Gs\rangle }{1152}\nonumber\\
&& -\frac{ig_s G^{a}_{\alpha\beta}t^a_{ij}(\!\not\!{x}
\sigma^{\alpha\beta}+\sigma^{\alpha\beta} \!\not\!{x})}{32\pi^2x^2}  -\frac{\delta_{ij}x^4\langle \bar{s}s \rangle\langle g_s^2 GG\rangle}{27648}-\frac{1}{8}\langle\bar{s}_j\sigma^{\mu\nu}s_i \rangle \sigma_{\mu\nu}   +\cdots \, ,
\end{eqnarray}
\begin{eqnarray}
C_{ij}(x)&=&\frac{i}{(2\pi)^4}\int d^4k e^{-ik \cdot x} \left\{
\frac{\delta_{ij}}{\!\not\!{k}-m_c}
-\frac{g_sG^n_{\alpha\beta}t^n_{ij}}{4}\frac{\sigma^{\alpha\beta}(\!\not\!{k}+m_c)+(\!\not\!{k}+m_c)
\sigma^{\alpha\beta}}{(k^2-m_c^2)^2}\right.\nonumber\\
&&\left. -\frac{g_s^2 (t^at^b)_{ij} G^a_{\alpha\beta}G^b_{\mu\nu}(f^{\alpha\beta\mu\nu}+f^{\alpha\mu\beta\nu}+f^{\alpha\mu\nu\beta}) }{4(k^2-m_c^2)^5}+\cdots\right\} \, ,\nonumber\\
f^{\lambda\alpha\beta}&=&(\!\not\!{k}+m_c)\gamma^\lambda(\!\not\!{k}+m_c)\gamma^\alpha(\!\not\!{k}+m_c)\gamma^\beta(\!\not\!{k}+m_c)\, ,\nonumber\\
f^{\alpha\beta\mu\nu}&=&(\!\not\!{k}+m_c)\gamma^\alpha(\!\not\!{k}+m_c)\gamma^\beta(\!\not\!{k}+m_c)\gamma^\mu(\!\not\!{k}+m_c)\gamma^\nu(\!\not\!{k}+m_c)\, ,
\end{eqnarray}
and  $t^n=\frac{\lambda^n}{2}$, the $\lambda^n$ is the Gell-Mann matrix \cite{Reinders85}, then compute  the integrals both in the coordinate and momentum spaces,  and obtain the correlation functions $\Pi_{\mu\nu}(p)$ therefore the spectral densities at the level of   quark-gluon degrees  of freedom. In Eq.(15), we retain the terms $\langle\bar{s}_j\sigma_{\mu\nu}s_i \rangle$  originate from the Fierz re-arrangement of the $\langle s_i \bar{s}_j\rangle$ to  absorb the gluons  emitted from other quark lines to  extract the mixed condensate $\langle\bar{s}g_s\sigma G s\rangle$ \cite{WangHuangtao-2014}.

 Once analytical expressions of the QCD spectral densities  are obtained,  we can take the
quark-hadron duality below the continuum thresholds  $s_0$ and perform Borel transform  with respect to
the variable $P^2=-p^2$ to obtain  the following four  QCD sum rules:
\begin{eqnarray}
\lambda^2_{Y}\, \exp\left(-\frac{M^2_{Y}}{T^2}\right)= \int_{4m_c^2}^{s_0} ds\, \rho(s) \, \exp\left(-\frac{s}{T^2}\right) \, ,
\end{eqnarray}
where $\rho(s)=\rho^1(s)$, $\rho^2(s)$, $\rho^3(s)$ and $\rho^4(x)$,
\begin{eqnarray}
\rho^1(s)&=&\rho_{0}(s)+\rho_{3}(s) +\rho_{4}(s)+\rho_{5}(s)+\rho_{6}(s)+\rho_{7}(s) +\rho_{8}(s)+\rho_{10}(s)\, ,
\end{eqnarray}
\begin{eqnarray}
\rho^2(s)&=&\rho^1(s)\mid_{m_s \to 0,\,\langle \bar{s}s\rangle \to \langle \bar{q}q\rangle, \,\langle \bar{s}g_s\sigma Gs\rangle \to \langle \bar{q}g_s\sigma G q\rangle } \, , \nonumber\\
\rho^3(s)&=&\rho^1(s)\mid_{m_c \to -m_c} \, , \nonumber\\
\rho^4(s)&=&\rho^3(s)\mid_{m_s \to 0,\,\langle \bar{s}s\rangle \to \langle \bar{q}q\rangle, \,\langle \bar{s}g_s\sigma Gs\rangle \to \langle \bar{q}g_s\sigma G q\rangle } \, ,
\end{eqnarray}
 the subscripts $i=0$, $3$, $4$, $5$, $6$, $7$, $8$, $10$ denote the dimensions of the vacuum condensates, the explicit expressions of the QCD  spectral densities $\rho_i(s)$ are presented in the Appendix.

 In this article, we carry out the operator product expansion to the vacuum condensates  up to dimension-10 and discard the  perturbative corrections, and assume vacuum saturation for the  higher dimensional  vacuum condensates.
The higher dimensional vacuum condensates  are always
 factorized to lower dimensional vacuum condensates with vacuum saturation in the QCD sum rules,
  factorization works well in  large $N_c$ limit.  In reality, $N_c=3$, some  (not much) ambiguities maybe come from
the vacuum saturation assumption. The higher dimensional vacuum condensates have not been well studied yet.

We derive   Eq.(17) with respect to  $\tau=\frac{1}{T^2}$, then eliminate the
 pole residues $\lambda_{Y}$, and obtain the QCD sum rules for
 the masses of the vector   tetraquark states,
 \begin{eqnarray}
 M^2_{Y}&=& -\frac{\int_{4m_c^2}^{s_0} ds\frac{d}{d \tau}\rho(s)\exp\left(-\tau s \right)}{\int_{4m_c^2}^{s_0} ds \rho(s)\exp\left(-\tau s\right)}\, .
\end{eqnarray}

\section{Numerical results and discussions}
We take  the standard values of the vacuum condensates $\langle
\bar{q}q \rangle=-(0.24\pm 0.01\, \rm{GeV})^3$,   $\langle
\bar{q}g_s\sigma G q \rangle=m_0^2\langle \bar{q}q \rangle$,
$m_0^2=(0.8 \pm 0.1)\,\rm{GeV}^2$, $\langle\bar{s}s \rangle=(0.8\pm0.1)\langle\bar{q}q \rangle$, $\langle\bar{s}g_s\sigma G s \rangle=m_0^2\langle \bar{s}s \rangle$,  $\langle \frac{\alpha_s
GG}{\pi}\rangle=(0.33\,\rm{GeV})^4 $    at the energy scale  $\mu=1\, \rm{GeV}$
\cite{SVZ79,Reinders85,Colangelo-Review}, and choose the $\overline{MS}$ masses $m_{c}(m_c)=(1.28\pm0.03)\,\rm{GeV}$, $m_s(\mu=2\,\rm{GeV})=0.096^{+0.008}_{-0.004}\,\rm{GeV}$ from the Particle Data Group \cite{PDG}, and set $m_u=m_d=0$.
Moreover, we take into account the energy-scale dependence of  the input parameters on the QCD side,
\begin{eqnarray}
\langle\bar{q}q \rangle(\mu)&=&\langle\bar{q}q \rangle(Q)\left[\frac{\alpha_{s}(Q)}{\alpha_{s}(\mu)}\right]^{\frac{12}{25}}\, , \nonumber\\
 \langle\bar{s}s \rangle(\mu)&=&\langle\bar{s}s \rangle(Q)\left[\frac{\alpha_{s}(Q)}{\alpha_{s}(\mu)}\right]^{\frac{12}{25}}\, , \nonumber\\
 \langle\bar{q}g_s \sigma Gq \rangle(\mu)&=&\langle\bar{q}g_s \sigma Gq \rangle(Q)\left[\frac{\alpha_{s}(Q)}{\alpha_{s}(\mu)}\right]^{\frac{2}{25}}\, , \nonumber\\ \langle\bar{s}g_s \sigma Gs \rangle(\mu)&=&\langle\bar{s}g_s \sigma Gs \rangle(Q)\left[\frac{\alpha_{s}(Q)}{\alpha_{s}(\mu)}\right]^{\frac{2}{25}}\, , \nonumber\\
m_c(\mu)&=&m_c(m_c)\left[\frac{\alpha_{s}(\mu)}{\alpha_{s}(m_c)}\right]^{\frac{12}{25}} \, ,\nonumber\\
m_s(\mu)&=&m_s({\rm 2GeV} )\left[\frac{\alpha_{s}(\mu)}{\alpha_{s}({\rm 2GeV})}\right]^{\frac{12}{25}} \, ,\nonumber\\
\alpha_s(\mu)&=&\frac{1}{b_0t}\left[1-\frac{b_1}{b_0^2}\frac{\log t}{t} +\frac{b_1^2(\log^2{t}-\log{t}-1)+b_0b_2}{b_0^4t^2}\right]\, ,
\end{eqnarray}
   where $t=\log \frac{\mu^2}{\Lambda^2}$, $b_0=\frac{33-2n_f}{12\pi}$, $b_1=\frac{153-19n_f}{24\pi^2}$, $b_2=\frac{2857-\frac{5033}{9}n_f+\frac{325}{27}n_f^2}{128\pi^3}$,  $\Lambda=210\,\rm{MeV}$, $292\,\rm{MeV}$  and  $332\,\rm{MeV}$ for the flavors  $n_f=5$, $4$ and $3$, respectively  \cite{PDG}, and evolve all the input parameters to the optimal energy scales   $\mu$ to extract the masses of the vector hidden-charm tetraquark states.

 In the present QCD sum rules, we search for the ideal  Borel parameters $T^2$ and continuum threshold
parameters $s_0$  to obey  the  following four criteria:\\
$\bf 1.$ Pole dominance at the phenomenological side;\\
$\bf 2.$ Convergence of the operator product expansion;\\
$\bf 3.$ Appearance of the Borel platforms;\\
$\bf 4.$ Satisfying the energy scale formula,\\
 using try and error.

 In Refs.\cite{ WangEPJC-4660,WangTetraquarkCTP,WangHuangtao-2014}, we study the energy scale dependence of the QCD sum rules in details
 and suggest an energy scale formula $\mu=\sqrt{M^2_{X/Y/Z}-(2{\mathbb{M}}_Q)^2}$   with the effective heavy quark mass ${\mathbb{M}}_Q$  to determine  the energy scales of the QCD spectral densities of the hidden-charm and hidden-bottom tetraquark states, which also works well for the hidden-charm pentaquark states \cite{WangPentaQuark}. In this article, we take the updated value ${\mathbb{M}}_c=1.82\,\rm{GeV}$ \cite{WangEPJC-1601}.

  The resulting Borel parameters or Borel windows $T^2$, continuum threshold parameters $s_0$, ideal energy scales of the QCD spectral densities, pole contributions of the ground state tetraquark states, and contributions of the vacuum condensates of dimension 7, 8 and 10 in the operator product expansion are shown   explicitly in Table 2. From the table, we can see that the first two criteria of the QCD sum rules are satisfied, so we expect to make reasonable predictions.

\begin{table}
\begin{center}
\begin{tabular}{|c|c|c|c|c|c|c|c|}\hline\hline
                       &$T^2 (\rm{GeV}^2)$ &$\sqrt{s_0}(\rm{GeV})$ &$\mu(\rm{GeV})$  &pole         &$|D_7|$        &$|D_8|$     &$|D_{10}|$\\ \hline

$c\bar{c}s\bar{s}^{1}$ &$3.6-4.0$          &$5.10\pm0.10$          &$2.9$            &$(41-60)\%$  &$< 1\% $       &$\ll 1\% $  &$\ll 1\% $   \\ \hline

$c\bar{c}q\bar{q}^{2}$ &$3.5-3.9$          &$5.05\pm0.10$          &$2.8$            &$(42-61)\%$  &$< 1\% $       &$\ll 1\% $  &$\ll 1\% $    \\ \hline

$c\bar{c}s\bar{s}^{3}$ &$3.2-3.6$          &$4.90\pm0.10$          &$2.6$            &$(40-60)\%$  &$< 1\% $       &$< 1\% $    &$\ll 1\% $    \\ \hline

$c\bar{c}q\bar{q}^{4}$ &$3.0-3.4$          &$4.75\pm0.10$          &$2.4$            &$(39-60)\%$  &$\sim 1\% $    &$< 1\% $    &$\ll 1\% $    \\ \hline
 \hline
\end{tabular}
\end{center}
\caption{ The Borel parameters, continuum threshold parameters, pole contributions, contributions of the vacuum condensates of dimension 7, 8 and 10, where the superscripts $1$, $2$, $3$ and $4$ denote the currents $J^1_\mu(x)$, $J^2_\mu(x)$,
$J^3_\mu(x)$ and $J^4_\mu(x)$, respectively. }
\end{table}

We take into account all uncertainties of the input parameters,
and obtain the values of the masses and pole residues of
 the   vector tetraquark states, which are  shown explicitly in Figs.1-2 and Table 3. From Figs.1-2, we can see that there appear platforms in the Borel windows, the  criterion  $\bf 3$ is  satisfied.
 From   Table 3, we can see that the  criterion  $\bf 4$ is also satisfied. Now the four criteria of the QCD sum rules are all satisfied, and we expect to make reliable predictions.

In Fig.1, we  also present the experimental values of the masses of the $Y(4260/4220)$, $Y(4360/4320)$, $Y(4390)$ and $Y(4660/4630)$ \cite{BES-Y4390,PDG}. From the figure, we can see that the experimental values of the masses $M_{Y(4660/4630)}$ and $M_{Y(4360/4320)}$ can be well reproduced,   the present predications support assigning the $Y(4660/4630)$  to be  the $C \otimes \gamma_\mu C$  type tetraquark state          $c\bar{c}s\bar{s}$, and assigning the $Y(4360/4320)$  to be the $C\gamma_5 \otimes \gamma_5\gamma_\mu C$  type tetraquark state  $c\bar{c}q\bar{q}$. The mass $M_{Y(4660/4630)}$ lies just below the upper bound of the predicted mass of the $C \otimes \gamma_\mu C$  type vector tetraquark state    $c\bar{c}q\bar{q}$, which disfavors assigning the $Y(4660/4630)$  to be the  $C \otimes \gamma_\mu C$  type vector tetraquark state  $c\bar{c}q\bar{q}$, however,  such an assignment  is not excluded.

If the relative P-waves between the diquark and antidiquark cost a
universal energy for all the tetraquark states, then the $C \otimes \gamma_\mu C$ type tetraquark states have larger masses than the corresponding $C\gamma_5 \otimes \gamma_5\gamma_\mu C$  type tetraquark states,
as
$C \otimes \gamma_\mu C=\left[C\gamma_5 \underline{\gamma_5} \otimes \gamma_\mu C\right]\oplus \left[C\gamma_\alpha \underline{\gamma^\alpha} \otimes \gamma_\mu C\right]$  and
$C\gamma_5 \otimes \gamma_5\gamma_\mu C=C\gamma_5 \otimes \underline{\gamma_5}\gamma_\mu C$, the $C\gamma_\mu$ diquark states have slightly  larger masses than the corresponding $C\gamma_5$ diquark states from the QCD sum rules \cite{WangDiquark,WangLDiquark}. In other words, the bad diquarks have slightly larger masses than  the good diquarks,  those  effects are also accounted for  in the effective Hamiltonian \cite{Maiani-II-type,Ali-Maiani-Y}.

 On the other hand, the $Y(4660)$ can be assigned to be the $\psi^\prime f_0(980)$ hadro-charmonium or molecular   state based on fitting the mass distribution of the process $e^+e^- \to \psi^\prime \pi^+\pi^-$ \cite{GuoFK-4660-psif0} or the calculations of the QCD sum rules \cite{WangZG-4660-psif0}. More experimental data are still needed to assign the $Y(4660)$ unambiguously.

In this article, we recalculate the QCD sides of the correlation functions for the $C \otimes \gamma_\mu C$ type currents by taking into account the neglected terms due to the approximations involving the higher  dimensional vacuum condensates in Ref.\cite{WangEPJC-4660} and correct a small error in numerical calculations, the present predictions   $4.66\pm0.09/4.59\pm0.08 \, \rm{GeV}$ are more robust than the values $4.70^{+0.14}_{-0.10}/4.66^{+0.17}_{-0.10}\,\rm{GeV}$ obtained in Ref.\cite{WangEPJC-4660}. In Fig.3, we plot the mass and pole residue of the $C\otimes \gamma_\mu C$ type vector tetraquark state $c\bar{c}s\bar{s}$ with variation of the Borel parameter $T^2$ for  truncations of the operator product expansion,  $D=6$, $7$, $8$ and $10$. From the figure, we can see that the higher dimensional  vacuum condensates play an important role in  determining the Borel platforms.

The ground state $C\gamma_5 \otimes \gamma_5C$ type  and
$C\gamma_\mu \otimes \gamma^\mu C$ type hidden-charm tetraquark states $c\bar{c}q\bar{q}$ have the masses about $3.85\,\rm{GeV}$ from the QCD sum rules  in which  the vacuum condensates up to dimension 10 are taken into account in a consistent way \cite{WangMPLA,WangTetraquarkCTP}, if an additional P-wave costs about $0.5\,\rm{GeV}$, the ground state vector  hidden-charm tetraquark states $c\bar{c}q\bar{q}$ have the mass about $4.35\,\rm{GeV}$, the present prediction $4.34\pm0.08\,\rm{GeV}$ is robust. In Ref.\cite{ZhangHuang-JHEP}, Zhang and Huang introduce an explicit P-wave in the currents, and obtain the value $4.32\pm0.20\,\rm{GeV}$ by taking into account the vacuum condensates up to dimension 6.

In Ref.\cite{Ali-Maiani-Y}, A. Ali et al study the hidden-charm P-wave tetraquarks and the newly excited charmed $\Omega_c$ states with the  effective Hamiltonian incorporating the dominant spin-spin, spin-orbit and tensor interactions,  and observe that the $Y(4220)$, $Y(4330)$, $Y(4390)$, $Y(4660)$ can be assigned
to be the four ground states with $L=1$ by fitting the coefficients in the effective Hamiltonian to the experimental masses. In the effective Hamiltonian approach,
the lowest state is the $Y(4220)$, while we cannot   obtain such low mass based on the QCD sum rules \cite{WangJPG,ZhangHuang-PRD,ZhangHuang-JHEP,Nielsen-4260-4460,WangEPJC-4660,WangEPJC-1601}.

The mass $M_{Y(4390)}$ lies just below the upper bound of the predicted mass of the $C\gamma_5 \otimes \gamma_5\gamma_\mu C$  type vector tetraquark state  $c\bar{c}q\bar{q}$, which disfavors assigning the $Y(4390)$  to be the  $C\gamma_5 \otimes \gamma_5\gamma_\mu C$  type vector tetraquark state  $c\bar{c}q\bar{q}$,
on the  other hand, the mass $M_{Y(4390)}$ lies just below the lower bound of the predicted mass of the $C\gamma_5 \otimes \gamma_5\gamma_\mu C$  type vector tetraquark state  $c\bar{c}s\bar{s}$, which also disfavors assigning the $Y(4390)$  to be the  $C\gamma_5 \otimes \gamma_5\gamma_\mu C$  type vector tetraquark state  $c\bar{c}s\bar{s}$,  however, such  assignments  are not completely excluded. If we take the energy scale $\mu=3.4\,\rm{GeV}$ for the $c\bar{c}s\bar{s}$ tetraquark state or $\mu=1.8\,\rm{GeV}$ for the $c\bar{c}q\bar{q}$ tetraquark state, the experimental value of the $M_{Y(4390)}$ can be reproduced, however, such energy scales are not consistent with the QCD sum rules for other tetraquark states.

 There are no candidates for the $Y(4260/4220)$ in the present calculations.  In Refs.\cite{Zhao-PRL-Y4260-3900,Zhao-PRL-Y4260-3900-2}, the $Y(4260)$ and $Z_c(3900)$ are assigned to be the $\bar{D}D_1(2420) +D\bar{D}_1(2420)$ and $\bar D D^*+D\bar D^*$ molecular states respectively based on the heavy
 meson (non-relativistic) effective field theory.
 In Ref.\cite{WangCPC-Y4390}, we study the vector molecular states $D\bar{D}_1(2420)$ and $D^*\bar{D}_0^*(2400)$ with the QCD sum rules
  by taking into account the vacuum condensates up to dimension-10 in the operator product expansion in a consistent way, and use the energy scale formula for the molecular states to determine the optimal energy scales of the QCD spectral densities \cite{WangHuang-molecule}, and  obtain the predications  $M_{D\bar{D}_1(1^{--})}=4.36\pm0.08\,\rm{GeV}$ and $M_{D^*\bar{D}_0^*(1^{--})}=4.78\pm0.07\,\rm{GeV}$.  The QCD sum rules support assigning the $Y(4390)$ (not the $Y(4260/4220)$) and $Z_c(3900)$ to be the $D\bar{D}_1$ and $D\bar{D}^*$ S-wave  molecular states, respectively \cite{WangCPC-Y4390,WangHuang-molecule}.
While the lattice QCD supports assigning the $Y(4260)$ to be a hybrid state \cite{Hybrid-4260-Lattice}.
Furthermore, there have been observed evidences for the  $X(3872)$  in the lattice calculations,  though its interpretation was not specified
 \cite{Lattice-X3872},  there was no evidence for the $Z_c(3900)$ in  the lattice calculations \cite{Lattice-Z3900}.
  There are also other assignments of the $Y(4390)$, for example, the  $D^*(2010)\bar{D}_1(2420)$ molecular state \cite{Hejun-4390}.

\begin{table}
\begin{center}
\begin{tabular}{|c|c|c|c|c|c|c|c|}\hline\hline
                        &$\mu(\rm{GeV})$    &$M_{Y}(\rm{GeV})$  &$\lambda_{Y}(10^{-2}\rm{GeV}^5)$ &Expt(MeV)    &Assignments       \\ \hline

$c\bar{c}s\bar{s}^{1}$  &$2.9$              &$4.66\pm0.09$      &$6.74\pm 0.88$                   &$4643\pm 9$  &$Y(4660)$       \\ \hline

$c\bar{c}q\bar{q}^{2}$  &$2.8$              &$4.59\pm0.08$      &$6.21\pm0.77$                    &             &       \\ \hline

$c\bar{c}s\bar{s}^{3}$  &$2.6$              &$4.49\pm0.09$      &$4.95\pm0.72$                    &             &            \\ \hline

$c\bar{c}q\bar{q}^{4}$  &$2.4$              &$4.34\pm0.08$      &$3.91\pm0.57$                    &$4341\pm 8$  &$Y(4360)$      \\ \hline
 \hline
\end{tabular}
\end{center}
\caption{ The  masses and pole residues of the vector tetraquark states, where the superscripts $1$, $2$, $3$ and $4$ denote the currents $J^1_\mu(x)$, $J^2_\mu(x)$,
$J^3_\mu(x)$ and $J^4_\mu(x)$, respectively. }
\end{table}

\begin{table}
\begin{center}
\begin{tabular}{|c|c|c|c|c|c|c|c|}\hline\hline
                 &$J_\mu^1(x)\, ,\mu=2.9\,\rm{GeV}$ &$J_\mu^1(x)\, ,\mu=2.4\,\rm{GeV}$ &$J_\mu^4(x)\, ,\mu=2.4\,\rm{GeV}$  &$J_\mu^4(x)\, ,\mu=2.9\,\rm{GeV}$ \\ \hline

$\lambda_{Y(4220)}$  &$0.38729$          &$0.29100$          &$0.02632$            &$1.56680$          \\ \hline

$\lambda_{Y(4320)}$  &$0.69720$          &$0.20867$          &$3.90360$            &$3.94290$          \\ \hline

$\lambda_{Y(4390)}$  &$0.41733$          &$0.41695$          &$0.00000$            &$0.00190$          \\ \hline

$\lambda_{Y(4660)}$  &$6.47460$          &$5.93670$          &$0.00000$            &$0.00000$          \\ \hline
 \hline
\end{tabular}
\end{center}
\caption{ The central values of the fitted   pole residues, where the unit is $10^{-2}\,\rm{GeV}^5$. }
\end{table}

\begin{figure}
 \centering
 \includegraphics[totalheight=5cm,width=7cm]{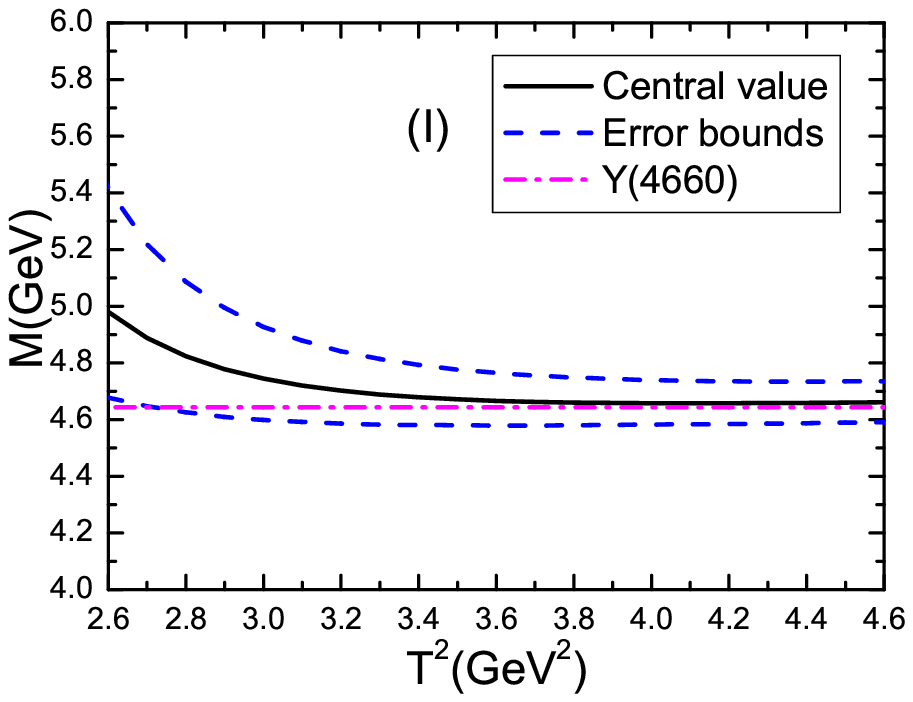}
 \includegraphics[totalheight=5cm,width=7cm]{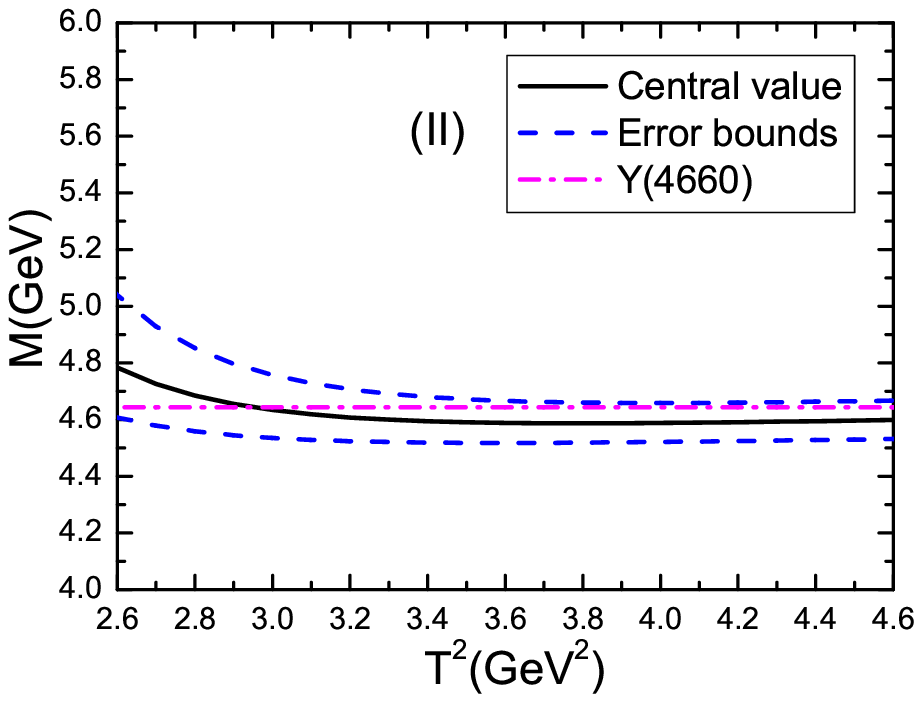}
  \includegraphics[totalheight=5cm,width=7cm]{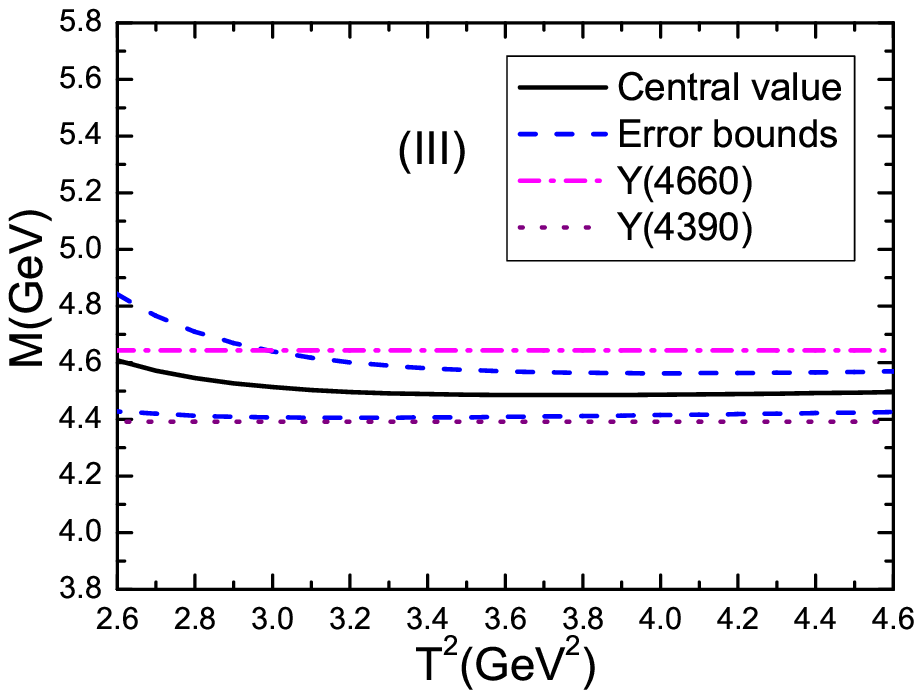}
 \includegraphics[totalheight=5cm,width=7cm]{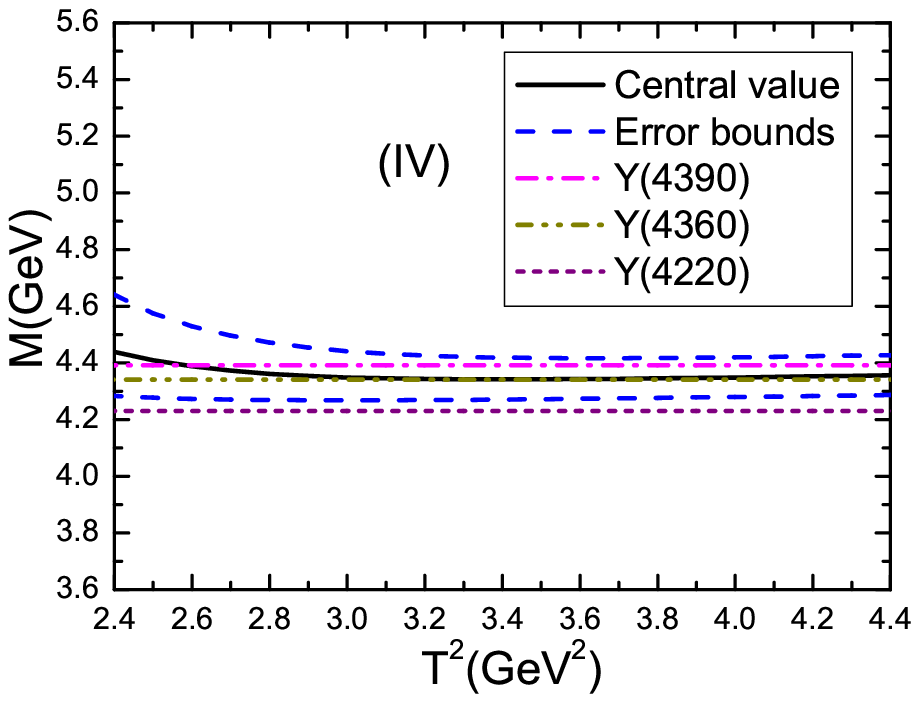}
         \caption{ The masses of the vector tetraquark states with variations of the Borel parameters $T^2$, where  the (I), (II), (III) and (IV) denote the currents $J^1_\mu(x)$, $J^2_\mu(x)$, $J^3_\mu(x)$ and $J^4_\mu(x)$, respectively.   }
\end{figure}

\begin{figure}
 \centering
 \includegraphics[totalheight=5cm,width=7cm]{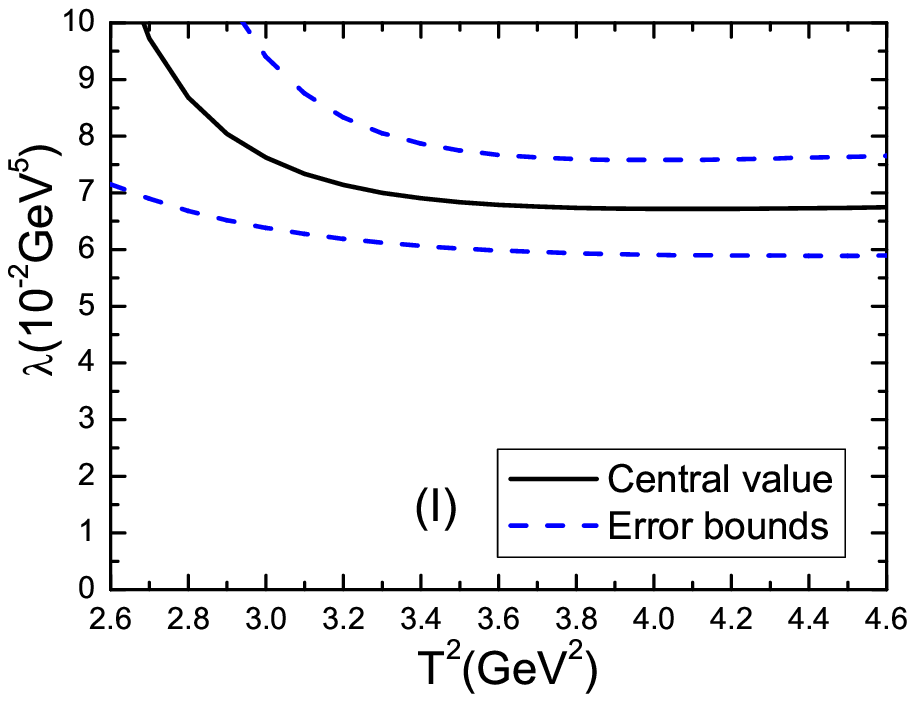}
 \includegraphics[totalheight=5cm,width=7cm]{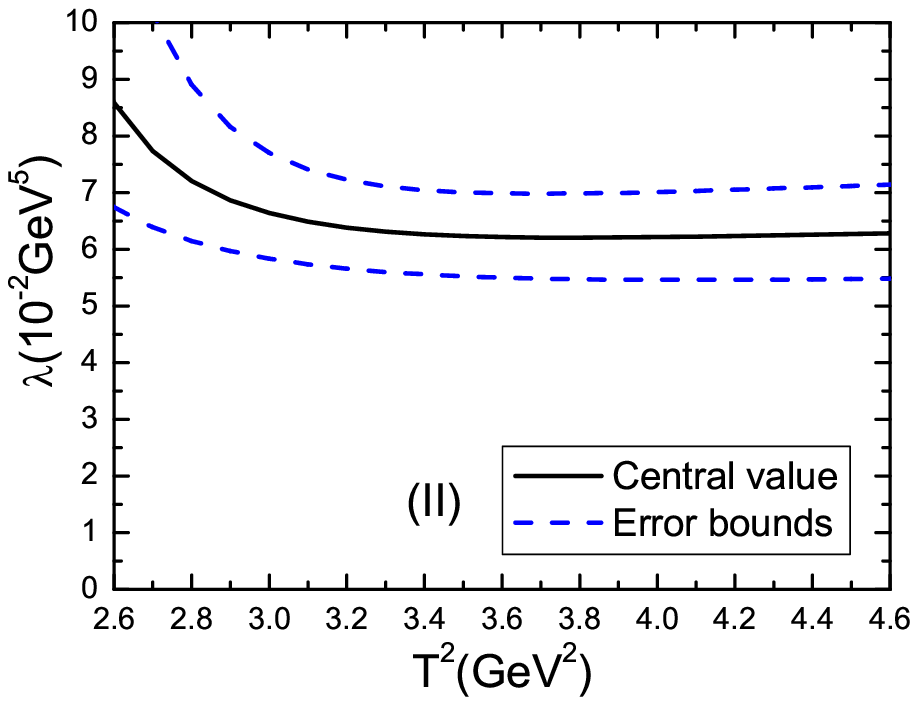}
  \includegraphics[totalheight=5cm,width=7cm]{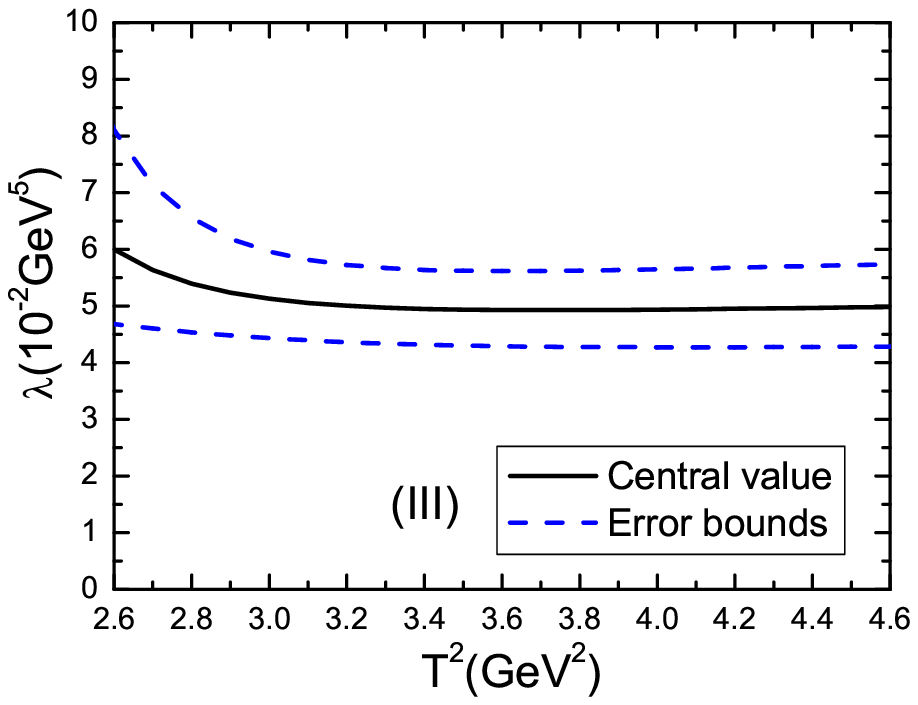}
 \includegraphics[totalheight=5cm,width=7cm]{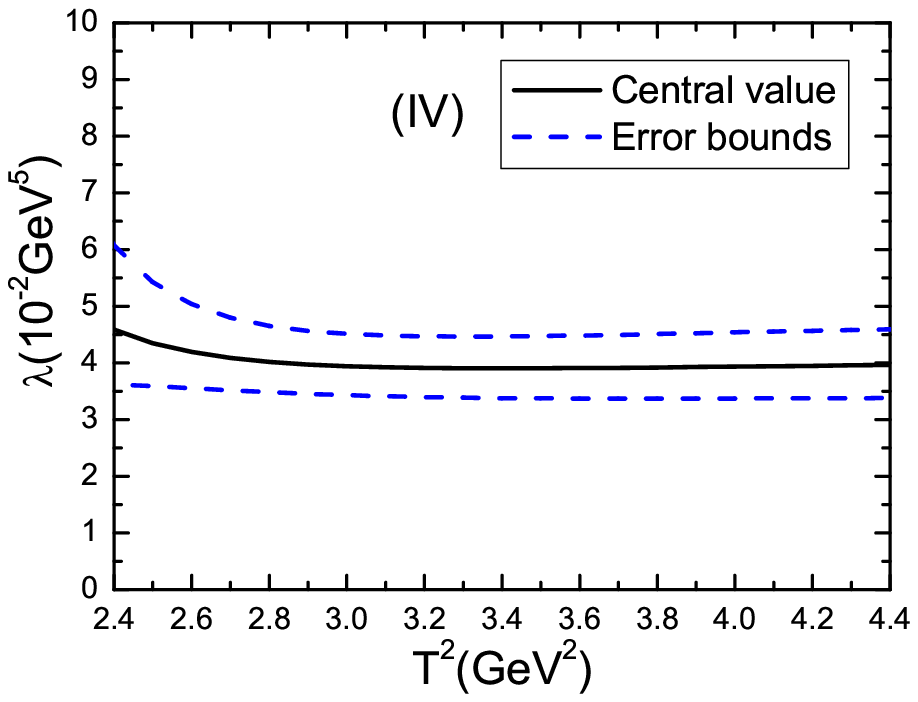}
         \caption{ The pole residues of the vector tetraquark states with variations of the Borel parameters $T^2$, where  the (I), (II), (III) and (IV) denote the currents $J^1_\mu(x)$, $J^2_\mu(x)$, $J^3_\mu(x)$ and $J^4_\mu(x)$, respectively.    }
\end{figure}

\begin{figure}
 \centering
 \includegraphics[totalheight=5cm,width=7cm]{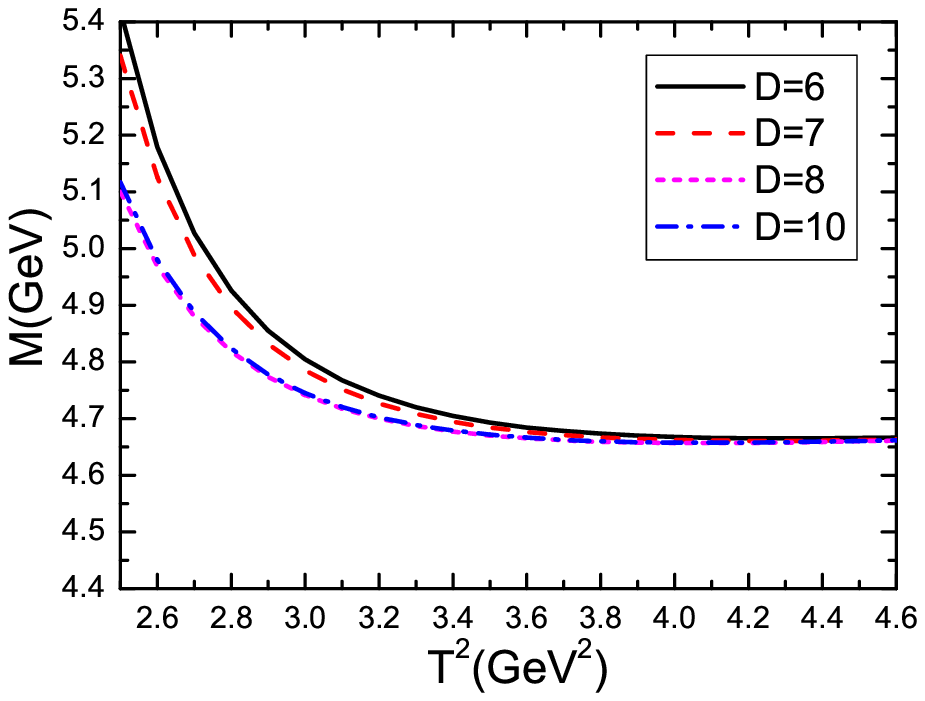}
 \includegraphics[totalheight=5cm,width=7cm]{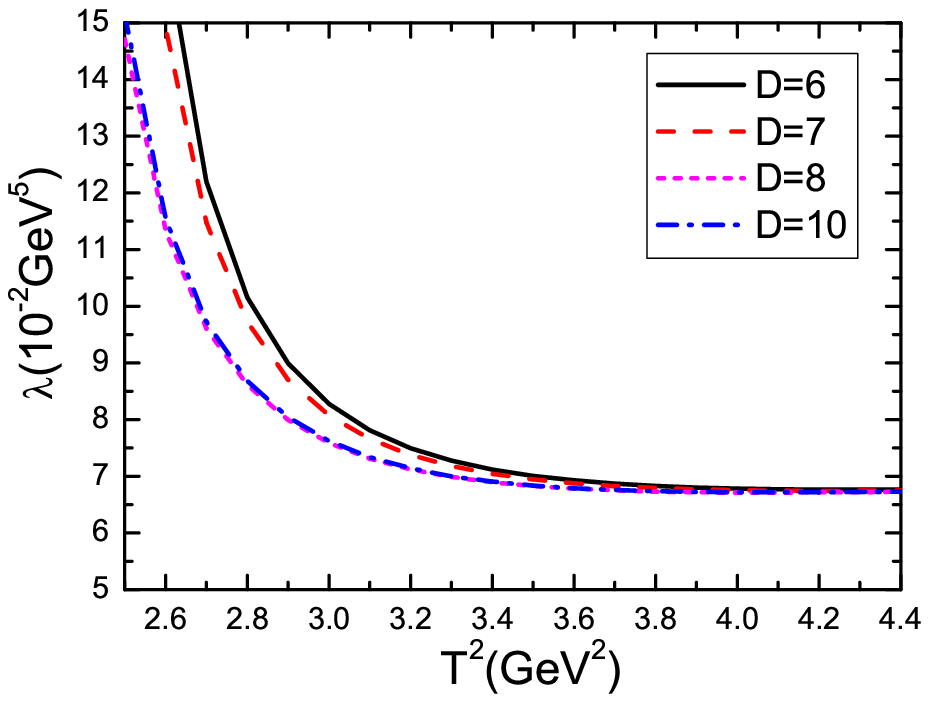}
         \caption{ The mass and pole residue of the $C\otimes \gamma_\mu C$ type vector tetraquark state $c\bar{c}s\bar{s}$ with variations  of the Borel parameter $T^2$, where  the  $D=6$, $7$, $8$ and $10$ denote  truncations of the operator product expansion.   }
\end{figure}

Now we check the assignment of the  $Y(4660/4630)$  as  the $C \otimes \gamma_\mu C$  type vector tetraquark state   $c\bar{c}s\bar{s}$ and the assignment of the $Y(4360/4320)$  as the $C\gamma_5 \otimes \gamma_5\gamma_\mu C$  type vector tetraquark state  $c\bar{c}q\bar{q}$ by assuming the currents $J^1_\mu(x)$ and $J^4_\mu(x)$ both couple  potentially to the four $Y$ states $Y(4260/4220)$, $Y(4360/4320)$, $Y(4390)$ and $Y(4660/4630)$. We take the experimental values $M_{Y(4220)}=4.230\,\rm{GeV}$, $M_{Y(4360)}=4.341\,\rm{GeV}$, $M_{Y(4390)}=4.392\,\rm{GeV}$ and $M_{Y(4660)}=4.643\,\rm{GeV}$ as input parameters \cite{BES-Y4390,PDG}, and take the pole residues $\lambda_Y$ as free parameters to fit the following two QCD sum rules,
\begin{eqnarray}
\Sigma_{Y=Y(4220), \,Y(4360), \,Y(4390), \, Y(4660)}\,\lambda^2_{Y}\, \exp\left(-\frac{M^2_{Y}}{T^2}\right)= \int_{4m_c^2}^{s_0} ds\, \rho^1(s) \, \exp\left(-\frac{s}{T^2}\right) \, ,
\end{eqnarray}
\begin{eqnarray}
\Sigma_{Y=Y(4220), \,Y(4360), \,Y(4390), \, Y(4660)}\,\lambda^2_{Y}\, \exp\left(-\frac{M^2_{Y}}{T^2}\right)= \int_{4m_c^2}^{s_0} ds\, \rho^4(s) \, \exp\left(-\frac{s}{T^2}\right) \, .
\end{eqnarray}

In Table 4, we present the central values of the fitted pole residues. In Fig.4, we plot the correlation functions with the central values of the fitted pole residues compared to the operator product expansion. From the figure, we can see that the QCD sides of the correlation functions can be well reproduced.   From   Table 4, we can see that the current $J^1_\mu(x)$ couples dominantly to the $Y(4660)$ both at the energy scales $\mu=2.9\,\rm{GeV}$ and $\mu=2.4\,\rm{GeV}$, the couplings to the $Y(4220)$, $Y(4360)$ and $Y(4390)$ can be neglected safely. For the current $J^4_\mu(x)$, if we take the ideal energy scale $\mu=2.4\,\rm{GeV}$, the   current $J^4_\mu(x)$ couples dominantly to the $Y(4360)$, the couplings to the $Y(4220)$, $Y(4390)$ and $Y(4660)$ can be neglected safely; on the other hand, if we take larger energy scale $\mu=2.9\,\rm{GeV}$, the   current $J^4_\mu(x)$ couples potentially to the $Y(4360)$, the coupling to the $Y(4220)$ is not neglectful.

In Refs.\cite{Ali-Maiani-Y,Faccini-4220}, the $Y(4220)$ is assigned to be the lowest vector tetraquark state in the simple diquark-antidiquark model with the constituent $c\bar{c}q\bar{q}$. In the present work, we can see that the lowest tetraquark states couple potentially to the current $J_\mu^4(x)$, not to the currents $J^1_\mu(x)$, $J^2_\mu(x)$ and $J^3_\mu(x)$, now we suppose that the $Y(4260/4220)$ is a pure vector tetraquark state and saturates  the QCD sum rules,
\begin{eqnarray}
 M^2_{Y(4260)}&=&- \frac{\int_{4m_c^2}^{s_0} ds\frac{d}{d \tau}\rho^4(s)\exp\left(-\tau s \right)}{\int_{4m_c^2}^{s_0} ds \rho^4(s)\exp\left(-\tau s\right)}\, ,
\end{eqnarray}
and study the energy scale dependence of the extracted mass $M_{Y(4260)} $ with the central values  of the input    parameters in Table 2.
In Fig.5, we plot the extracted mass $M_{Y(4260)}$ with variations of the Borel parameters $T^2$ and energy scales $\mu$. From the figure, we can see that the
mass $M_{Y(4260)}$ decreases  monotonously with increase of the energy scales $\mu$, the platforms appear at about $T^2= 3\,\rm{GeV}^2$. Even at the large energy scale $\mu=5\,\rm{GeV}$, the extracted mass $M_{Y(4260)}> 4.230\,\rm{GeV}$, so the $Y(4260/4220)$ is unlikely to be a pure vector tetraquark state.

If we take the energy scale formula $\mu=\sqrt{M^2_{X/Y/Z}-(2{\mathbb{M}}_c)^2}$ with the effective mass ${\mathbb{M}}_c=1.82\,\rm{GeV}$ as a constraint, the QCD sum rules only support assigning the $Y(4660/4630)$ and $Y(4360/4320)$ to be the  $C \otimes \gamma_\mu C$ type vector tetraquark $c\bar{c}s\bar{s}$ and $C\gamma_5 \otimes \gamma_5\gamma_\mu C$  type vector tetraquark state $c\bar{c}q\bar{q}$ respectively, and
  disfavor  assigning the $Y(4660/4630)$ (or $Y(4390)$) to be the  $C \otimes \gamma_\mu C$ (or $C\gamma_5 \otimes \gamma_5\gamma_\mu C$)  type vector tetraquark state  $c\bar{c}q\bar{q}$.

\begin{figure}
 \centering
 \includegraphics[totalheight=5cm,width=7cm]{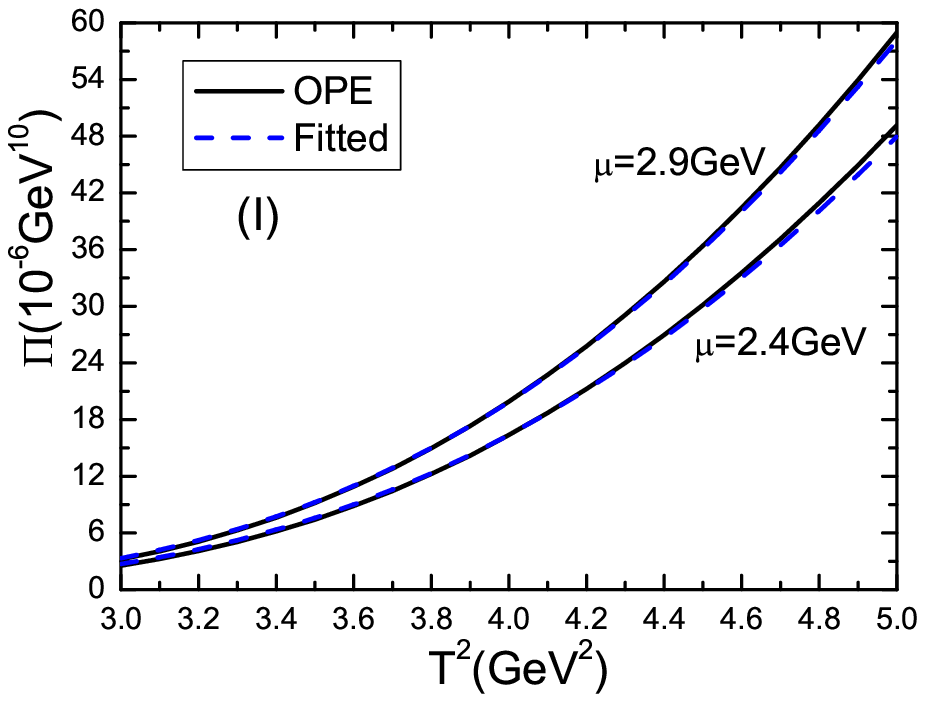}
 \includegraphics[totalheight=5cm,width=7cm]{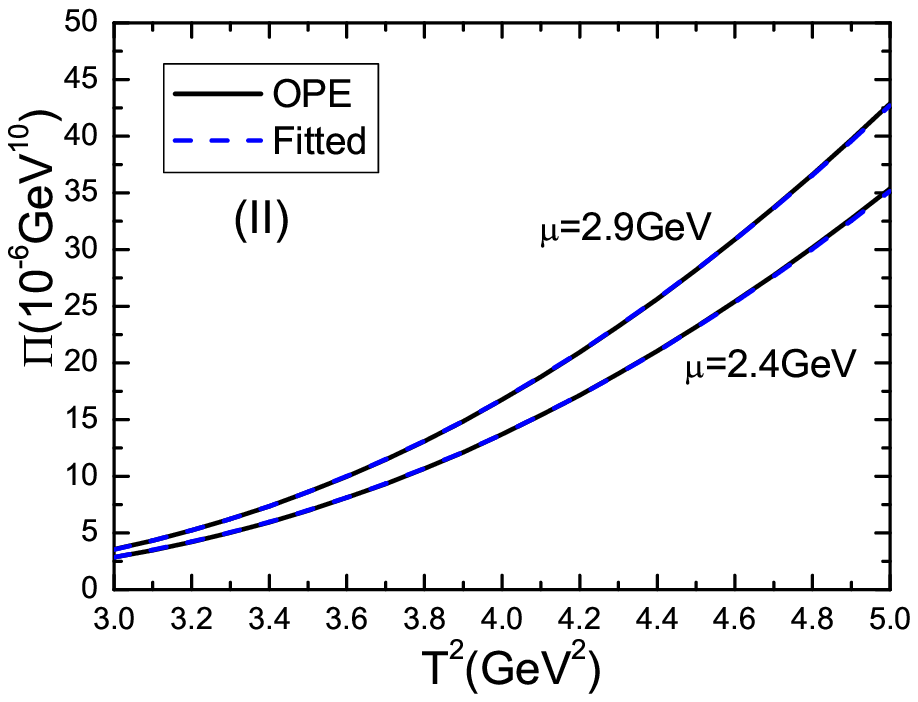}
         \caption{ The correlation functions with the central values of the fitted pole residues compared to the operator product expansion (OPE), where  the (I) and (II) denote the currents $J^1_\mu(x)$ and $J^4_\mu(x)$, respectively.   }
\end{figure}

\begin{figure}
 \centering
   \includegraphics[totalheight=5cm,width=7cm]{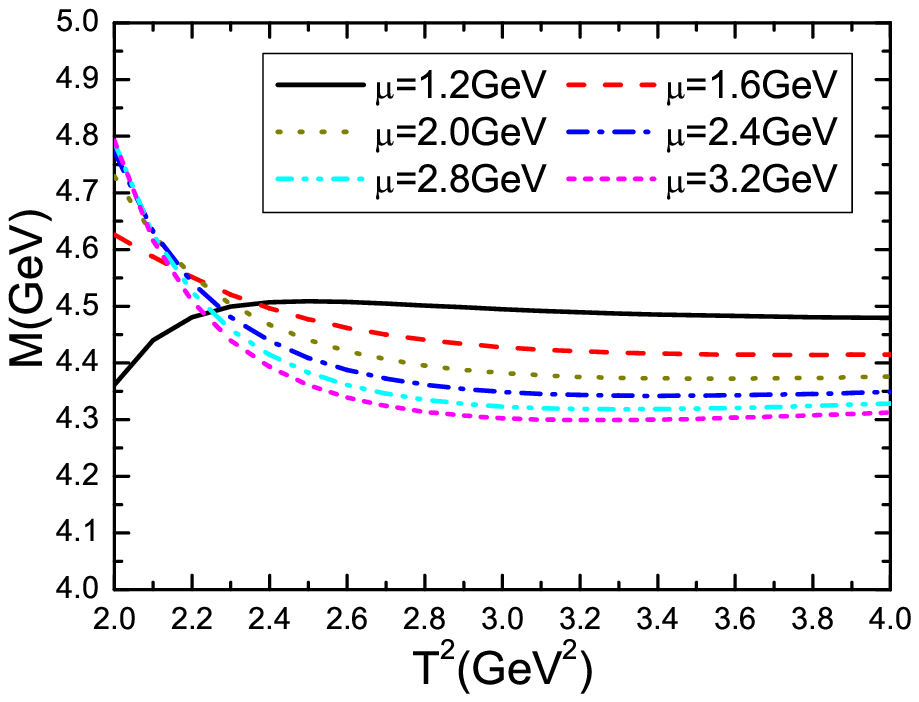}
 \includegraphics[totalheight=5cm,width=7cm]{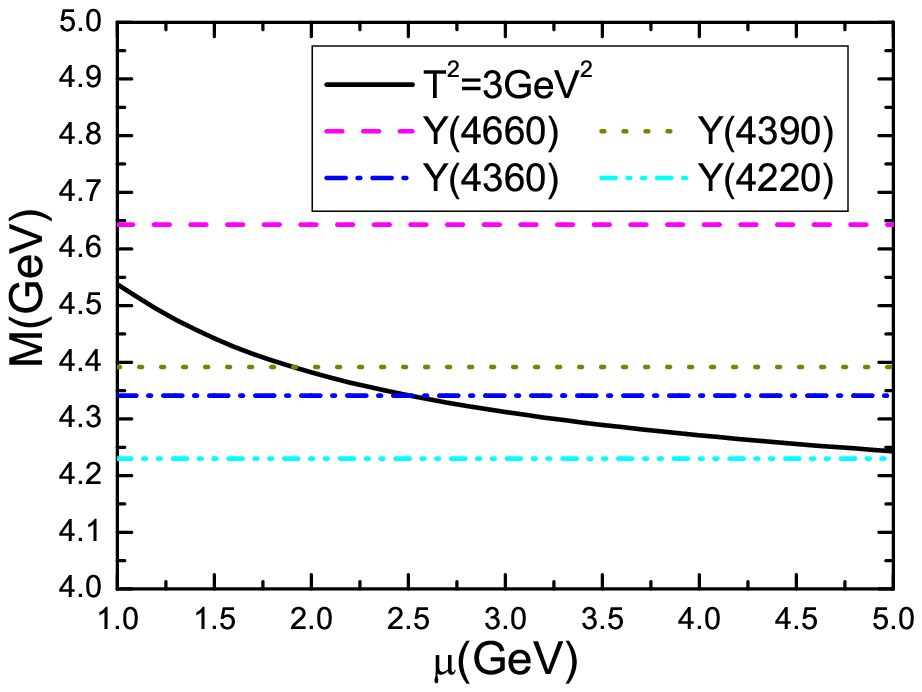}
         \caption{ The extracted mass with variations of the Borel parameters $T^2$ and energy scales $\mu$.   }
\end{figure}

\section{Conclusion}
In this article, we construct the $C \otimes \gamma_\mu C$ and $C\gamma_5 \otimes \gamma_5\gamma_\mu C$ type  currents to interpolate  the vector   tetraquark states,  then calculate the contributions of the vacuum condensates up to dimension-10  in the operator product expansion in a consistent way, and obtain four QCD sum rules. In calculations,  we use the  formula $\mu=\sqrt{M^2_{X/Y/Z}-(2{\mathbb{M}}_c)^2}$ to determine  the optimal energy scales of the QCD spectral densities, explore the energy scale dependence of the QCD sum rules in details, moreover, we take the experimental values of the masses of the $Y(4260/4220)$, $Y(4360/4320)$, $Y(4390)$ and $Y(4660/4630)$ as input parameters and  fit the pole residues to reproduce the correlation functions at the QCD side. The numerical results support assigning the $Y(4660/4630)$  to be the  $C \otimes \gamma_\mu C$ type vector tetraquark state $c\bar{c}s\bar{s}$, assigning the $Y(4360/4320)$ to be $C\gamma_5 \otimes \gamma_5\gamma_\mu C$  type vector tetraquark state $c\bar{c}q\bar{q}$,  and disfavor assigning the $Y(4260/4220)$ and $Y(4390)$ to be the pure vector tetraquark states.

\section*{Appendix}

The QCD spectral densities $\rho_0(s)$, $\rho_3(s)$, $\rho_5(s)$, $\rho_6(s)$, $\rho_7(s)$, $\rho_8(s)$ and  $\rho_{10}(s)$,
\begin{eqnarray}
\rho_{0}(s)&=&\frac{1}{1024\pi^6}\int dydz \, yz(1-y-z)^2\left(s-\overline{m}_c^2\right)^3\left(5s-\overline{m}_c^2 \right)  \nonumber\\
&&+\frac{m_c^2}{1536\pi^6}\int dydz \, (1-y-z)^3\left(s-\overline{m}_c^2\right)^3   \, ,
\end{eqnarray}
\begin{eqnarray}
\rho_{3}(s)&=&\frac{m_s\langle \bar{s}s\rangle}{16\pi^4}\int dydz \, yz\, s\left(s-\overline{m}_c^2\right) +\frac{m_sm_c^2\langle \bar{s}s\rangle}{32\pi^4}\int dydz \,(5-y-z) \left(s-\overline{m}_c^2\right)\, ,
\end{eqnarray}

\begin{eqnarray}
\rho_{4}(s)&=&-\frac{m_c^2}{768\pi^4} \langle\frac{\alpha_s GG}{\pi}\rangle\int dydz \left( \frac{z}{y^2}+\frac{y}{z^2}\right)(1-y-z)^2 \left( 2s-\overline{m}_c^2\right) \nonumber\\
&&-\frac{m_c^4}{4608\pi^4} \langle\frac{\alpha_s GG}{\pi}\rangle\int dydz \left( \frac{1}{y^3}+\frac{1}{z^3}\right)(1-y-z)^3   \nonumber\\
&&+\frac{m_c^2}{1536\pi^4} \langle\frac{\alpha_s GG}{\pi}\rangle\int dydz \left( \frac{1}{y^2}+\frac{1}{z^2}\right)(1-y-z)^3 \left( s-\overline{m}_c^2\right) \nonumber\\
&&+\frac{1}{768\pi^4}\langle\frac{\alpha_s GG}{\pi}\rangle\int dydz \, (y+z)   (1-y-z)\,  s\left(s-\overline{m}_c^2 \right) \nonumber\\
&&+\frac{1}{1024\pi^4}\langle\frac{\alpha_s GG}{\pi}\rangle\int dydz \,    (1-y-z)^2 \,\overline{m}_c^2\, \left(s-\overline{m}_c^2 \right) \nonumber\\
&&-\frac{m_c^2}{18432\pi^4}\langle\frac{\alpha_s GG}{\pi}\rangle\int dydz \, \frac{(1-y-z)^2(5+y+z)}{yz}    \left(s-\overline{m}_c^2 \right) \, ,
\end{eqnarray}

\begin{eqnarray}
\rho_5(s)&=&-\frac{m_s\langle \bar{s}g_s\sigma Gs\rangle}{192\pi^4}\int dy  \, y(1-y) \left(s+7\widetilde{m}_c^2 \right) \nonumber\\
&&+\frac{m_s m_c^2\langle \bar{s}g_s\sigma Gs\rangle}{128\pi^4}\int dydz  \, \left(\frac{3}{y}+\frac{3}{z}-2\right)  \nonumber\\
&&+\frac{m_c\langle \bar{s}g_s\sigma Gs\rangle}{128\pi^4}\int dydz  \, \left(\frac{z}{y}+\frac{y}{z}\right)(1-y-z) \left(2s-\overline{m}_c^2 \right) \nonumber\\
&&+\frac{m_c\langle \bar{s}g_s\sigma Gs\rangle}{256\pi^4}\int dydz  \,  (2-3y-3z) \left(s-\overline{m}_c^2 \right) \nonumber\\
&&-\frac{m_c\langle \bar{s}g_s\sigma Gs\rangle}{1536\pi^4}\int dydz  \, \left(\frac{z}{y}+\frac{y}{z}\right)(1-y-z) \left(5s-3\overline{m}_c^2 \right)\, ,
\end{eqnarray}

\begin{eqnarray}
\rho_6(s)&=&-\frac{m_c^2\langle\bar{s}s\rangle^2}{12\pi^2}\int dy+ \frac{ \langle\bar{s}s\rangle^2}{24\pi^2}\int dy\, y(1-y)\left(s-\widetilde{m}_c^2 \right)\, ,
\end{eqnarray}

\begin{eqnarray}
\rho_7(s)&=&-\frac{m_c\langle\bar{s}s\rangle}{64\pi^2}\langle\frac{\alpha_sGG}{\pi}\rangle\int dydz \left\{1+\frac{4}{9}\,s\,\delta\left(s-\overline{m}_c^2 \right)\right\} \nonumber\\
&&+\frac{m_c\langle\bar{s}s\rangle}{384\pi^2}\langle\frac{\alpha_sGG}{\pi}\rangle\int dydz \left(\frac{y}{z}+\frac{z}{y} \right) \, ,
\end{eqnarray}

\begin{eqnarray}
\rho_8(s)&=&\frac{m_c^2\langle\bar{s}s\rangle\langle\bar{s}g_s\sigma Gs\rangle}{24\pi^2}\int_0^1 dy \left(1+\frac{s}{T^2} \right)\delta\left(s-\widetilde{m}_c^2\right)\nonumber\\
&&-\frac{\langle\bar{s}s\rangle\langle\bar{s}g_s\sigma Gs\rangle}{16\pi^2}\int_0^1 dy\,y(1-y) \left\{1+\frac{s}{3}\,\delta\left(s-\widetilde{m}_c^2\right)\right\} \nonumber\\
&&+\frac{\langle\bar{s}s\rangle\langle\bar{s}g_s\sigma Gs\rangle}{192\pi^2}\int_{y_i}^{y_f} dy   \left\{1-2s\,\delta\left(s-\widetilde{m}_c^2\right)\right\} \nonumber\\
&&+\frac{m_s m_c\langle\bar{s}s\rangle\langle\bar{s}g_s\sigma Gs\rangle}{192\pi^2  }\int_0^1 dy\,  \delta\left(s-\widetilde{m}_c^2\right)  \nonumber\\
&&-\frac{m_s m_c\langle\bar{s}s\rangle\langle\bar{s}g_s\sigma Gs\rangle}{2304\pi^2  }\int_0^1 dy\, \left( \frac{1-y}{y}+\frac{y}{1-y}\right) \left(1-10\frac{s}{T^2}\right)\,\delta\left(s-\widetilde{m}_c^2\right)  \, ,
\end{eqnarray}

\begin{eqnarray}
\rho_{10}(s)&=&\frac{\langle\bar{s}g_s\sigma Gs\rangle^2}{64\pi^2 }\int_0^1 dy \,y(1-y) \left( 1+\frac{2s}{3T^2}+\frac{s^2}{6T^4}-\frac{s^3}{3T^6} \right)\, \delta \left( s-\widetilde{m}_c^2\right)
\nonumber\\
&&+\frac{m_c^4\langle\bar{s}s\rangle^2}{216T^4}\langle\frac{\alpha_sGG}{\pi}\rangle\int_0^1 dy  \left\{ \frac{1}{y^3}+\frac{1}{(1-y)^3}\right\} \delta\left( s-\widetilde{m}_c^2\right)\nonumber\\
&&-\frac{m_c^2\langle\bar{s}s\rangle^2}{72T^2}\langle\frac{\alpha_sGG}{\pi}\rangle\int_0^1 dy  \left\{ \frac{1}{y^2}+\frac{1}{(1-y)^2}\right\} \delta\left( s-\widetilde{m}_c^2\right)\nonumber\\
&&-\frac{m_c^2\langle\bar{s}s\rangle^2}{432T^2}\langle\frac{\alpha_sGG}{\pi}\rangle\int_0^1 dy  \left\{ \frac{1-y}{y^2}+\frac{y}{(1-y)^2}\right\} \delta\left( s-\widetilde{m}_c^2\right)\nonumber\\
&&-\frac{\langle\bar{s}g_s\sigma Gs\rangle^2}{384\pi^2 }\int_0^1 dy \, \left( 1+\frac{s}{2T^2} -\frac{s^2}{T^4}\right)\, \delta \left( s-\widetilde{m}_c^2\right)
\nonumber\\
&&+\frac{35\langle\bar{s}g_s\sigma Gs\rangle^2}{36864\pi^2  }\int_0^1 dy   \, \delta \left( s-\widetilde{m}_c^2\right)
\nonumber\\
&&+\frac{m_s m_c\langle\bar{s}g_s\sigma Gs\rangle^2}{1152\pi^2 T^2 }\int_0^1 dy \left( \frac{1-y}{y}+\frac{y}{1-y}\right) \left( 1+\frac{s}{T^2}-\frac{s^2}{T^4} \right)\, \delta \left( s-\widetilde{m}_c^2\right)
\nonumber\\
&&-\frac{m_s m_c\langle\bar{s}g_s\sigma Gs\rangle^2}{1152\pi^2 T^2 }\int_0^1 dy   \left( 1+\frac{s}{T^2}  \right)\, \delta \left( s-\widetilde{m}_c^2\right)
\nonumber\\
&&-\frac{m_s m_c\langle\bar{s}g_s\sigma Gs\rangle^2}{13824\pi^2 T^2 }\int_0^1 dy \left( \frac{1-y}{y}+\frac{y}{1-y}\right) \left( 1+\frac{s}{T^2}-\frac{2s^2}{T^4} \right)\, \delta \left( s-\widetilde{m}_c^2\right)
\nonumber\\
&&+\frac{\langle\bar{s}s\rangle^2}{72  } \langle\frac{\alpha_sGG}{\pi}\rangle\int_0^1 dy \,y(1-y)  \left(1+\frac{2s}{3T^2}+\frac{s^2}{6T^4}-\frac{s^3}{3T^6}\right) \, \delta\left( s-\widetilde{m}_c^2\right) \, ,
\end{eqnarray}
 where $\int dydz=\int_{y_i}^{y_f}dy \int_{z_i}^{1-y}dz$, $y_{f}=\frac{1+\sqrt{1-4m_c^2/s}}{2}$,
$y_{i}=\frac{1-\sqrt{1-4m_c^2/s}}{2}$, $z_{i}=\frac{y
m_c^2}{y s -m_c^2}$, $\overline{m}_c^2=\frac{(y+z)m_c^2}{yz}$,
$ \widetilde{m}_c^2=\frac{m_c^2}{y(1-y)}$, $\int_{y_i}^{y_f}dy \to \int_{0}^{1}dy$, $\int_{z_i}^{1-y}dz \to \int_{0}^{1-y}dz$,  when the $\delta$ functions $\delta\left(s-\overline{m}_c^2\right)$ and $\delta\left(s-\widetilde{m}_c^2\right)$ appear.

\section*{Acknowledgements}
This  work is supported by National Natural Science Foundation, Grant Number  11775079.

\end{document}